\PassOptionsToPackage{table}{xcolor}
\documentclass[5p,twocolumn]{elsarticle}

\usepackage{amsmath,amsfonts}
\usepackage{algorithmic}
\usepackage{algorithm}
\usepackage{array}
\usepackage{subfig}
\usepackage{url}
\usepackage{verbatim}
\usepackage{graphicx}
\usepackage{makecell}
\usepackage{xcolor}
\usepackage{enumitem}
\usepackage{paralist}
\usepackage{listings}
\usepackage{tabularx}
\usepackage{multirow}
\usepackage{arydshln}
\usepackage{float}
\usepackage{amssymb}
\usepackage{lineno} %

\usepackage{tikz}
\usetikzlibrary{positioning, fit, shapes.geometric, arrows.meta, calc, backgrounds}

\newlist{inlinelist}{enumerate*}{1}
\setlist[inlinelist,1]{label=(\roman*)}

\newlist{inlinenumber}{enumerate*}{1}
\setlist[inlinenumber,1]{label=(\arabic*)}

\newcommand{\fig}[1]{Fig.~\ref{fig:#1}}
\newcommand{\tab}[1]{Table~\ref{tab:#1}} \newcommand{\s}[1]{Sec.~\ref{sec:#1}}

\newcommand{\algo}[1]{Algorithm~\ref{algo:#1}}

\newcommand{\fakeparagraph}[1]{\smallskip\noindent\textbf{#1.}}

\newif\ifhidenotes
\hidenotesfalse

\ifhidenotes
  \newcommand{\noteam}[1]{}
  \newcommand{\notedd}[1]{}
  \newcommand{\notead}[1]{}
  \newcommand{\notemc}[1]{}
\else
  \newcommand{\noteam}[1]{\footnote{\color{blue}{Ale: #1}}}
  \newcommand{\notead}[1]{\footnote{\color{red}{Arianna: #1}}}
  \newcommand{\notedd}[1]{\footnote{\color{purple}{Dario: #1}}}
\fi

\newcommand\T{\rule{0pt}{2.6ex}}       %
\newcommand\B{\rule[-1.2ex]{0pt}{0pt}} %

\definecolor{numb}{rgb}{0.58,0,0.82}
\definecolor{punct}{rgb}{0.3,0.3,0.3}
\definecolor{delim}{rgb}{0.6,0.6,0.6}

\usepackage{hyperref}

\journal{Future Generation Computer Systems}

\begin{document}

\begin{frontmatter}

  \title{Ermes: a Stateful Serverless Platform for the Edge-to-Cloud Continuum}

  \author[polimi]{Matteo Cenzato}
  \ead{matteo.cenzato@polimi.it}

  \author[polimi]{Dario d'Abate\corref{cor1}}
  \ead{dario.dabate@polimi.it}

  \author[polimi]{Arianna Dragoni\corref{cor1}}
  \ead{arianna.dragoni@polimi.it}

  \author[polimi]{Giacomo Orsenigo}
  \ead{giacomo.orsenigo@mail.polimi.it}

  \author[polimi]{Luca Tosetti}
  \ead{luca.tosetti@mail.polimi.it}

  \author[polimi]{Matteo Briscini}
  \ead{matteo.briscini@mail.polimi.it}

  \author[polimi]{Alessandro Margara}
  \ead{alessandro.margara@polimi.it}
  \cortext[cor1]{Corresponding author}

  \affiliation[polimi]{organization={Politecnico di Milano}, city={Milan},
    country={Italy}}

\begin{abstract}
  Function-as-a-Service (FaaS) is a widely adopted paradigm to simplify
  application deployment across the edge-to-cloud continuum. However, its
  stateless nature forces functions to retrieve their state from external,
  typically cloud-centric, data stores, reintroducing the very latency that edge
  computing aims to eliminate.
  This issue is further exacerbated by location-agnostic schedulers and rigid,
  one-size-fits-all consistency models that fail to capture the diverse
  requirements of edge applications.
  In this paper, we propose Ermes, a distributed platform that natively
  integrates state management into the FaaS paradigm, enabling the joint
  distribution of computational workloads and application state across the
  edge-to-cloud continuum.
  Ermes organizes application state into logical units, termed
  \emph{collections}, and employs a distributed coordination algorithm that
  jointly maps collections and functions onto the available nodes seeking to
  minimize the latency perceived by the clients.
  In addition, it supports fine-grained replication and per-collection
  consistency levels, ranging from sequential to eventual consistency, leaving
  developers the choice of how to resolve the trade-off between consistency and
  performance.
  The experimental evaluation shows that Ermes quickly turns remote state
  accesses into local ones and sustains low latency as the workload grows, and
  as clients move across the edge.
\end{abstract} 

\end{frontmatter}

\sloppy

\section{Introduction}
\label{sec:introduction}
Cloud computing has long served as the backbone of distributed applications,
offering virtually unlimited computation and storage resources~\cite{armbrust-cloud}.
However, the physical distance between clients and remote data centers
introduces unavoidable latency, which becomes prohibitive for a growing class of
latency-sensitive applications~\cite{shi-promise}.

Edge computing addresses this limitation by distributing resources across
small-scale nodes located near the end-users~\cite{shi-challenges}.
The edge-to-cloud continuum integrates these decentralized resources with
traditional cloud services~\cite{bonomi-fog}, creating a unified infrastructure
where applications can dynamically balance workloads between the cloud and the
network periphery.

In this landscape, serverless computing, and specifically
Function-as-a-Service (FaaS), has emerged as a widely adopted paradigm to
simplify application deployment~\cite{Serverless_computing_state_of_the_art}.
Under the FaaS model, the execution logic is decoupled from infrastructure
management. Indeed, the platform handles resource provisioning and scaling
automatically, which is particularly beneficial in the edge-to-cloud
continuum, where resources are inherently heterogeneous and geographically
distributed.

However, to facilitate the dynamic placement and migration of workloads, FaaS
functions are typically designed to be stateless~\cite{serverlessStepBack},
which keeps the model lightweight and portable and enables rapid scaling even at
the constrained network periphery.
Within this model, the application state resides in external storage, typically
centralized in the cloud. As a consequence, edge-deployed functions must fetch
their state from remote data stores, reintroducing the very latency that edge
computing was intended to eliminate.

A first step toward mitigating this problem is to account for locality when
placing computation. Several edge-oriented FaaS platforms already do so, routing
invocations to nodes close to the requesting client~\cite{neptune, serverledge,
  dfaas, rausch-skippy, cicconetti-decentralized}.
These platforms, however, remain stateless: they optimize the placement of
computation while leaving the application state in external storage.
As a result, a function scheduled close to its client may still incur a remote
round-trip to fetch its state. Location-aware scheduling alone is therefore
insufficient when the state itself is not brought close to the client.
To address this, a more effective strategy involves co-locating application
state and functions at the network periphery.
However, optimizing state placement is non-trivial, as multiple functions
often require access to shared state portions.
Replicating the state across edge nodes closer to the clients can reduce
access latency, but it introduces a fundamental trade-off, as keeping replicas
consistent requires coordination, and the cost of this coordination depends
on the guarantees required.
Strict consistency demands synchronization that may offset the latency benefits
of edge placement, while weaker models reduce overhead at the expense of
allowing temporary divergence across replicas.

Most existing platforms do not move state toward the edge at all, and the few
that do address this complexity only in part. Some bring a single copy of the
state close to the client without replicating it, so that any other client
accessing it still pays a remote round-trip. Others do replicate, but enforce
rigid, one-size-fits-all consistency models~\cite{capTheorem} that lack the
flexibility to support the diverse requirements of edge applications.
As a result, the fundamental trade-off between data consistency and the
low-latency performance required at the edge remains largely unresolved in
current serverless architectures like Enoki~\cite{enoki}, Apache Flink StateFun\footnote{\url{https://github.com/apache/flink-statefun}}, and Faasm~\cite{faasm}.

In this paper, we propose Ermes, a distributed architecture that natively
integrates state management into the FaaS paradigm, enabling the joint
distribution of both computational workloads and application state across the
edge-to-cloud continuum.
Ermes organizes application state into logical units termed \emph{collections},
which are dynamically placed among available nodes to ensure that state is
persisted and accessed in proximity to where computation occurs.
To optimize placement, the framework employs a dual-engine approach.
It distributes collections in proximity to the end-users based on both locality
and resource availability, while simultaneously steering function execution
toward the nodes hosting the relevant state, or as close as possible to them.
To this end, Ermes automatically maps functions and state onto the available
nodes through a distributed coordination algorithm that seeks to minimize the
latency perceived by the clients.
This coordination algorithm was proposed and validated in a companion
work~\cite{ermesTheory}, which formulates the joint placement and scheduling
problem and shows, in simulation, that its decentralized heuristic closely
approximates an optimal allocation of the available resources

To complement this decentralized storage model, Ermes supports fine-grained
collections replication at the network periphery, supporting the diverse
requirements of modern edge applications.
Indeed, Ermes directly addresses the limitations of rigid, one-size-fits-all
consistency models empowering developers to resolve the trade-off between
consistency and performance by defining per-collection consistency levels.
In doing so, it leaves developers the choice of whether to prioritize strict
consistency or lower access latency, ranging from sequential consistency for
sensitive data to eventual consistency for performance-critical tasks.

We integrate within Ermes a set of decentralized protocols that run among the nodes
with no central coordinator, each deciding where to execute an invocation and
where to keep its collections from the demand it observes.

Whereas our companion work~\cite{ermesTheory} studied the placement and
scheduling heuristic in simulation, here we evaluate the fully implemented
platform on real infrastructure, assessing its performance and scalability
across the edge-to-cloud continuum.
Furthermore, we compared our solution against a traditional centralized baseline
to quantify the advantages of our decentralized approach.
The results demonstrate that Ermes delivers significant performance benefits
over conventional FaaS architectures, particularly in terms of request latency
and system throughput, validating the framework as a robust solution that
effectively bridges the gap between serverless simplicity and the performance
requirements of modern edge applications.
The remainder of this paper is organized as follows.
We position Ermes within the landscape of serverless and edge computing in
\s{related-works}.
We then present its programming model (\s{prog_model}) and an overview of the
platform (\s{sys_overview}), before detailing its two families of
functionalities, function management (\s{func:mng}) and state management
(\s{state:mng}).
We describe our implementation in \s{implementation} and report the experimental
evaluation in \s{eval}.
Finally, \s{conclusion} concludes the paper and outlines directions for future
work.

\section{Related Work}
\label{sec:related-works}

This section positions Ermes within the landscape of serverless and edge
computing, which we survey along three lines of research. The first develops
lightweight execution \emph{runtimes}, focused on running individual functions
on a single node. The second builds full-fledged \emph{platforms} that
coordinate function execution across multiple nodes in the edge-to-cloud
continuum. The third studies function scheduling and data placement as an
optimization problem, in isolation from any specific system. We examine the
three in turn, and close by positioning Ermes with respect to all of them.

\fakeparagraph{Runtimes}
Lightweight execution runtimes minimize invocation overhead, making them well
suited for resource-constrained edge devices.
Runtimes such as \emph{Sledge}~\cite{sledge}, \emph{TinyFaaS}~\cite{tinyFaas},
\emph{Faasm}~\cite{faasm}, and \emph{Faasd} are designed as lightweight
execution engines for resource-constrained edge nodes, but none of them
coordinates work across multiple nodes.
Most adopt a stateless model, where functions have no persistent state across
invocations. The exception is \emph{Faasm}, which maintains persistent state
and replicates it across function instances for fault tolerance.
However, these runtimes are not full-fledged platforms: they do not include
coordination protocols to manage function scheduling and state placement
across geographically distributed nodes.

Our own runtime, \emph{WASP}~\cite{wasp}, belongs to this same category but is
designed to be modular: it decouples function execution from state management,
turning the execution engine, the datastore, and the caching policies into
pluggable components that administrators tailor to each node without altering
application code. Like the others, WASP runs on a single node and does not
coordinate execution across the hierarchy, a role that Ermes fulfills by building
on it (\s{func:mng}).

\fakeparagraph{Platforms}
Commercial cloud FaaS platforms, managed by public vendors, address the
scheduling problem by assuming homogeneous and virtually infinite resources,
transparently instantiating new function instances without exposing placement
decisions to the developer.
Moreover, they follow a stateless execution model, delegating all data
persistence to external storage services.
The most widely adopted (\emph{AWS Lambda}\footnote{\url{https://aws.amazon.com/lambda}}, \emph{Google Cloud Functions}\footnote{\url{https://cloud.google.com/functions}}, and
\emph{Azure Functions}\footnote{\url{https://azure.microsoft.com/products/functions}}) follow a purely stateless model, where functions
retain no persistent state across invocations, and any data persistence must
be handled explicitly through external storage services.

Some commercial platforms extend this model with native state management.
\emph{Azure Durable Functions} introduces \emph{Entity Functions} that expose
persistent state as if it was local memory, although the underlying storage
remains centralized in a single region.
\emph{Cloudflare Durable Objects}\footnote{\url{https://developers.cloudflare.com/durable-objects/}} couples computation with storage following
the actor model, binding each object to a single location close to its first
client request to guarantee strong consistency, but without support for
offloading or replication across nodes.
\emph{AWS IoT Greengrass}\footnote{\url{https://aws.amazon.com/greengrass}} extends the FaaS paradigm to IoT devices at the far
edge, providing state abstraction through the Device Shadow service and
automatically replicating local state to the cloud for durability.
However, all these platforms are designed to run on the vendor's own
infrastructure, with no support for deployment on private, user-managed nodes.
Moreover, most of them operate exclusively within cloud datacenters.

The two exceptions that target edge scenarios are Cloudflare Durable Objects
and Greengrass. However, they still tie execution to vendor-controlled
deployment models.
Indeed, Cloudflare Durable Objects execute on the vendor's globally
distributed edge nodes, not on user-owned infrastructure, while Greengrass
runs on user-owned IoT devices but is orchestrated entirely from the cloud
console.

\newcolumntype{C}[1]{>{\centering\arraybackslash}m{#1}}
\begin{table*}[t]
  \scriptsize
  \centering
  \begin{tabular}{|C{10em} | C{5.5em} | C{5.5em} | C{5.5em} | C{5.5em} |
      C{5.5em}| C{5.5em} | C{5.5em} | C{5.5em} | C{5.5em}|}
    \hline
    \rowcolor{gray!50}             & \textbf{Decentralized}     &
    \textbf{State model}           & \textbf{State Abstraction} & \textbf{Scheduling
                                                                    Policy}     & \textbf{Locality Awareness} & \textbf{Resource Awareness} &
    \textbf{Data Replication}\T\B                                                                                                             \\
    \hline
    \textbf{OpenWhisk}             & $\times$                   & Stateless
                                   & -                          & Centralized            & $\times$
                                   & \checkmark                 & - \B
    \\
    \hline
    \textbf{Lean OpenWhisk}        & $\times$                   & Stateless
                                   & -                          & Centralized            & $\times$
                                   & \checkmark                 & - \B
    \\
    \hline
    \textbf{OpenFaaS}              & $\times$                   & Stateless
                                   & -                          & Centralized            & $\times$
                                   & \checkmark                 & - \B
    \\
    \hline
    \textbf{Funless}               & $\times$                   & Stateless
                                   & -                          & Centralized            & $\times$
                                   & \checkmark                 & - \B
    \\
    \hline
    \textbf{Neptune}               & \checkmark                 & Stateless
                                   & -                          & Centralized            & \checkmark
                                   & \checkmark                 & - \B
    \\
    \hline
    \textbf{Edgeless}              & \checkmark                 & Stateless
                                   & -                          & Hybrid                 & $\times$
                                   & \checkmark                 & - \B
    \\
    \hline
    \textbf{DFaaS}                 & \checkmark                 & Stateless
                                   & -                          & Distributed            & $\times$
                                   & \checkmark                 & - \B
    \\
    \hline
    \textbf{Serverledge}           & \checkmark                 & Stateless
                                   & -                          & Distributed            & \checkmark
                                   & \checkmark                 & - \B
    \\
    \hline
    \textbf{Cloudburst}            & $\times$                   & Stateful
                                   & \checkmark                 & Centralized            & $\times$
                                   & \checkmark                 & \checkmark \B
    \\
    \hline
    \textbf{Apache Flink Statefun} & $\times$                   & Stateful
                                   & \checkmark                 & Centralized            & $\times$
                                   & \checkmark                 & \checkmark \B
    \\
    \hline
    \textbf{Enoki}                 & \checkmark                 & Stateful
                                   & \checkmark                 & Trivial
                                   & \checkmark                 & $\times$               &
    \checkmark \B
    \\
    \hline
    \textbf{StructMesh}            & \checkmark                 & Stateful
                                   & $\times$                   & Centralized            & $\times$
                                   & \checkmark                 & $\times$ \B
    \\
    \hline
    \textbf{FaDO}                  & \checkmark                 & Stateful
                                   & $\times$                   & Centralized            & \checkmark
                                   & \checkmark                 & \checkmark \B
    \\
    \hline
    \textbf{Ermes}                 & \textbf{\checkmark}        & \textbf{Stateful}
                                   & $\checkmark$               & \textbf{Distributed}   & \textbf{\checkmark}
                                   & \textbf{\checkmark}        & \textbf{\checkmark} \B
    \\
    \hline
  \end{tabular}
  \caption{State model and scheduling policy overview}
  \label{tab:related}
\end{table*}

Open-source and academic platforms remove this constraint, as they can be
deployed on arbitrary, user-managed infrastructures.
We summarize the most representative solutions in \tab{related}, comparing
them along relevant dimensions.
Specifically, we distinguish platforms that natively coordinate function
execution across geographically distributed nodes (i.e., \emph{decentralized}
platforms) from those that confine operations to a single cluster or region.
We then distinguish between \emph{stateful} platforms that manage persistent
state across invocations and \emph{stateless} ones that delegate persistence
to external services.
Within the former, we evaluate whether the platform provides high-level
primitives to shield developers from underlying storage complexities
(\emph{state abstraction}).
Finally, we examine how each platform makes function placement decisions.
We distinguish platforms based on their decision-making logic. In
\emph{centralized} platforms, a global scheduler dictates task allocation,
whereas \emph{distributed} platforms rely on autonomous peer-to-peer
negotiation. We also identify \emph{trivial} placement, where execution is
performed by the node receiving the client request without forwarding, and
\emph{hybrid} approaches, which partition the infrastructure into active
schedulers and passive executors.
We also evaluate whether the policy accounts for proximity to the requesting
client (\emph{locality awareness}) and available capacity on candidate nodes
(\emph{resource awareness}), and whether the platform supports \emph{data
  replication} across nodes.

\emph{Apache OpenWhisk}~\cite{openWhisk}, \emph{Lean
  OpenWhisk}\footnote{\url{https://github.com/kpavel/incubator-openwhisk/tree/lean}}, \emph{OpenFaaS}~\cite{openFaas}, and
\emph{FunLess}~\cite{funless} deploy a central control plane that manages
all scheduling decisions based on available node capacity.
Moreover, orchestration is confined to a single cluster with no support for
geo-distributed coordination.
\emph{Neptune}~\cite{neptune} extends coordination beyond a single cluster,
supporting placement across geographically distributed nodes. Indeed, its
hierarchical scheduler accounts for both resource availability and client
proximity, but placement decisions are still delegated to a central
coordinator.
\emph{Edgeless}~\cite{edgeless} takes a hybrid scheduling approach, where a
designated subset of nodes collectively manages scheduling for the rest based
on available resources, but without accounting for client proximity.
\emph{DFaaS}~\cite{dfaas} and \emph{Serverledge}~\cite{serverledge} adopt a
fully distributed scheduling policy, where nodes autonomously decide whether
to handle a request locally or forward it to a neighbor. \emph{DFaaS} bases
this decision purely on load, while \emph{Serverledge} also accounts for
network proximity to the client.

All the aforementioned platforms follow a stateless model, delegating data
persistence to external services.
A different class of platforms addresses this limitation by natively
integrating a persistence layer into the serverless architecture.
\emph{Enoki}~\cite{enoki} adopts a local-first strategy. It provides state
abstraction through keygroups and proactively replicates data to the executing
node upon invocation, ensuring locality awareness.
However, its scheduling is trivial, and it does not account for resource
availability.

\emph{Cloudburst} and \emph{Apache Flink StateFun}\footnote{\url{https://github.com/apache/flink-statefun}}
provide both state abstraction and data replication, with a centralized
scheduler that accounts for resource availability. They are designed for
single-datacenter deployments and do not support geo-distributed
orchestration.
\emph{Enoki}~\cite{enoki}supports geo-distributed orchestration and provides
state abstraction. However, its scheduling is trivial, as execution is bound
to the receiving node with no possibility of offloading, and it does not
account for resource availability.
\emph{StructMesh}~\cite{structMesh} also adopts a decentralized architecture,
but relies on a centralized scheduler that considers only resource
availability. Moreover, unlike the previous platforms, it does not support
data replication.
\emph{FaDO}~\cite{fado} combines a decentralized architecture with data
replication and a centralized scheduler that accounts for both resource
availability and client proximity. However, like \emph{StructMesh}, it does
not provide state abstraction, exposing raw storage interfaces to the
developer.

To summarize, none of the existing platforms combines a decentralized stateful
model with state abstraction, a distributed, locality-aware scheduling policy,
and native data replication with configurable per-collection consistency.

\fakeparagraph{Scheduling and data placement}
A distinct line of research studies function scheduling and data placement as an
optimization problem in isolation, independently of any specific system.
Some works schedule functions without modeling their data dependencies, deciding
where to execute each invocation based on load, network proximity, or
energy~\cite{cicconetti-decentralized, rausch-skippy}.
Others jointly optimize function scheduling and data placement, but either leave
replication out of the model or treat consistency coarsely, without
distinguishing among different consistency levels.
Our companion work~\cite{ermesTheory} addresses precisely this gap, formulating
the joint placement and scheduling problem under heterogeneous consistency
models and deriving a decentralized heuristic that closely approximates the
optimum.

Ermes draws on all these lines of work. It is a full-fledged platform that
coordinates function execution across the hierarchy, executing functions on top
of the WASP runtime and realizing the decentralized heuristic of our companion
work~\cite{ermesTheory} to jointly place computation and state.
In doing so, it combines the lightweight, configurable execution of edge
runtimes, the multi-node coordination of platforms, and the locality- and
consistency-aware placement studied by scheduling algorithms, a combination that
no existing system offers. %
\section{Programming Model}
\label{sec:prog_model}
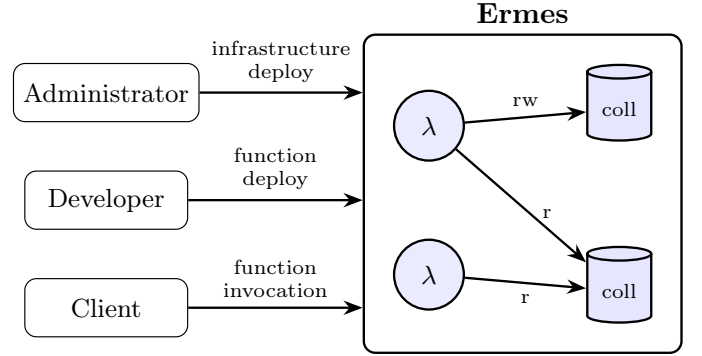
\begin{figure}[tpb]
  \centering
  \resizebox{\columnwidth}{!}{%
    \begin{tikzpicture}[
        font=\small, >=Stealth,
        func/.style={circle, draw, thick, fill=blue!8, minimum size=2.4em},
        coll/.style={draw, thick, cylinder, shape border rotate=90, aspect=0.25,
            fill=blue!10, minimum width=2.2em, minimum height=2.6em,
            font=\scriptsize},
        actor/.style={draw, rounded corners, minimum width=5.6em, minimum height=2em,
            align=center},
        albl/.style={midway, above, font=\scriptsize, align=center},
      ]

      \node[func] (f1) at (0,0)    {$\lambda$};
      \node[func] (f2) at (0,-1.8) {$\lambda$};

      \node[coll] (c1) at (2.3,0.2)  {coll};
      \node[coll] (c2) at (2.3,-2.0) {coll};

      \draw[->, thick] (f1) -- node[above, font=\scriptsize] {rw} (c1);
      \draw[->, thick] (f2) -- node[below, font=\scriptsize] {r}  (c2);
      \draw[->, thick] (f1) -- node[pos=0.7, above, font=\scriptsize] {r} (c2);

      \node[draw, thick, rounded corners, fit=(f1)(f2)(c1)(c2),
        inner sep=1em, label={[font=\bfseries]above:Ermes}] (ermes) {};

      \node[actor] (admin)  at (-3.9,0.4)  {Administrator};
      \node[actor] (dev)    at (-3.9,-0.9) {Developer};
      \node[actor] (client) at (-3.9,-2.2) {Client};

      \draw[->, thick] (admin)  -- node[albl] {infrastructure\\deploy}
      (admin  -| ermes.west);
      \draw[->, thick] (dev)    -- node[albl] {function\\deploy}
      (dev    -| ermes.west);
      \draw[->, thick] (client) -- node[albl] {function\\invocation}
      (client -| ermes.west);

    \end{tikzpicture}}
  \caption{External actors interacting with Ermes.}
  \label{fig:ermes-black-box}
\end{figure}

Ermes exposes its functionalities to three types of external actors, whose
interactions with the platform are illustrated in \fig{ermes-black-box}.
An \emph{administrator} provisions and deploys the underlying infrastructure
on which the platform runs. \emph{Developers} implement the application logic
as \emph{functions} and deploy them to the platform. \emph{Clients} trigger
the execution of these functions through invocation requests, and the
functions access the application state on their behalf.

Ermes adopts the Function-as-a-Service paradigm, but, unlike traditional FaaS
platforms, it is inherently stateful: functions can create, access, and modify
persistent application state during their execution.
The application state is organized into \emph{collections}, the unit of state
that functions create and operate on. Each collection is a set of
\emph{records}, key-value pairs that functions read and write individually.
Clients are limited in accessing collections. Each collection carries an
\emph{access policy}, fixed when it is created, that determines who may operate
on it: a \emph{private} collection is accessible only to its \emph{owner}, the
client on whose behalf it was created; a \emph{public} collection to every
client; and a \emph{shared} collection to a restricted set of clients. To
express sharing, developers organize clients into \emph{groups}: the creator of
a shared collection names the groups authorized to access it, and access is
granted to the owner and to any client that belongs to at least one of them.
\emph{Functions} are the fundamental units of computation in Ermes, and the
only means through which clients interact with the application state.
When a client invokes a function, it provides a set of arguments, and the
function executes on the client's behalf, operating exclusively on the
collections the client is authorized to access.
A function manipulates the state through a small, uniform set of operations.
It reads and writes the individual records of a collection through
\texttt{get} and \texttt{put} operations, and it is the sole mechanism for
creating new collections, fixing their consistency policy at creation time,
and for deleting existing ones.
Ermes maintains two kinds of metadata. A function is deployed with
\emph{function metadata}: it declares the collections it accesses by their
properties rather than by their identity, with placeholders that the
invocation's arguments fill in, so a function never names its collections
directly. Each collection, in turn, carries \emph{collection metadata}: its
properties, identity, owner, and consistency and access policies. At
invocation time, the platform resolves the function metadata against the
collection metadata, matching the properties a function declares against those
each collection carries, so that the same function operates on different
collections depending on the invoking client and the arguments it supplies.
Any record a function writes, and any collection it creates, persists beyond
the invocation, so that subsequent invocations, even of different functions
and on behalf of different clients, observe it according to the consistency
policy of the collection.

As a running example, consider a patient monitoring application deployed in a
hospital. A developer implements and deploys two functions:
\texttt{detectAnomaly}, which analyzes a patient's vitals to identify health
anomalies, and \texttt{analyzePopulation}, which examines the medical history
of a group of patients to detect common patterns. Each patient is a client
whose monitoring device invokes \texttt{detectAnomaly}, while doctors invoke
\texttt{analyzePopulation} over the data of the patients they follow.

Ermes replicates collections, and thus some consistency guarantees should be
preserved. To do so, Ermes provides two different consistency policies for a
collection.
A \emph{sequentially consistent} collection ensures that all the read and
write operations on the collection appear to the clients as if they were
executed in a sequential order, and that the operations of each client appear
in the order in which the client issued them. Developers thus obtain
serializable accesses and read-your-writes, even when a client moves and its
invocations are served by a different node.
An \emph{eventually consistent} collection relaxes ordering in favor of
latency. A client is guaranteed that the responses it observes eventually
converge to a single state, but has no guarantee during the transient, where
it may read different states, and concurrent updates to the same record are
not all preserved.
In the patient monitoring example, a developer stores the clinical data used
to detect anomalies in a sequentially consistent collection, to preserve
causality over life-critical records, and the append-only audit log of the
accesses in an eventually consistent one, to favor throughput over ordering.

A single invocation can access at most one sequentially consistent collection,
and an arbitrary number of eventually consistent ones. This restriction is
what allows Ermes to enforce a total order without running a consensus
protocol: with a single collection involved, the updates of an invocation are
ordered by construction, whereas spanning multiple sequentially consistent
collections would require a distributed commit, whose stop-the-world barrier
is prohibitive when inter-node latencies span the edge-to-cloud continuum.
This restriction is far less limiting than it may appear, as it mirrors a
design choice common to distributed databases that support multiple
consistency levels and deliberately confine strong guarantees to a single unit
of data to avoid global coordination bottlenecks. Apache Cassandra, for
instance, provides lightweight transactions that guarantee linearizability
only within a single partition, with no cross-partition
support\footnote{\url{https://cassandra.apache.org}}.
Ermes adopts the same principle: confining each invocation to a single
sequentially consistent collection removes cross-collection coordination
entirely, while still leaving eventually consistent collections unrestricted.

\section{System Overview}
\label{sec:sys_overview}

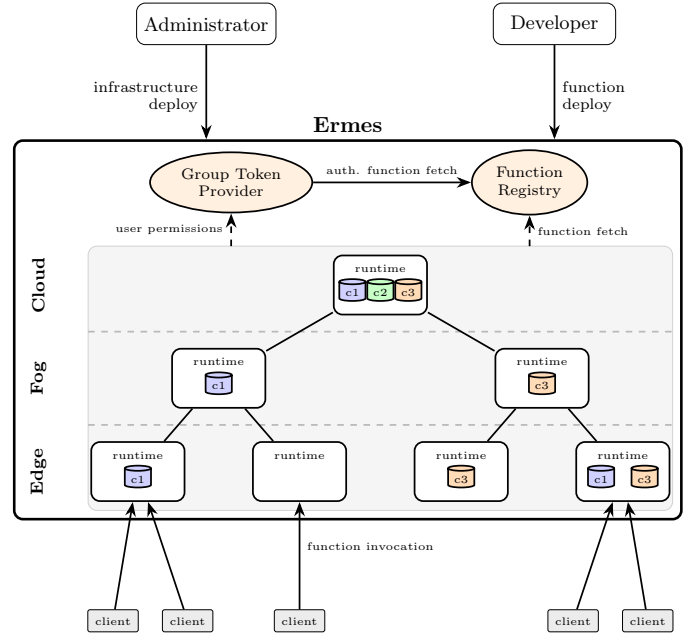
\begin{figure}[tpb]
  \centering
  \resizebox{\columnwidth}{!}{%
    \begin{tikzpicture}[
        >=Stealth, font=\small,
        rt/.style={draw, thick, rounded corners, fill=white,
            minimum width=1.5cm, minimum height=0.95cm},
        coll/.style={draw, thick, cylinder, shape border rotate=90, aspect=0.28,
            minimum width=1.2em, minimum height=1.0em, inner sep=1pt, font=\tiny},
        cA/.style={coll, fill=blue!18}, cB/.style={coll, fill=green!22},
        cC/.style={coll, fill=orange!28},
        cc/.style={draw, thick, ellipse, fill=orange!12, align=center,
            minimum width=5em, minimum height=2.2em, font=\scriptsize},
        cli/.style={draw, rounded corners=1pt, fill=gray!15,
            minimum width=1.5em, minimum height=1.0em, font=\tiny},
        actor/.style={draw, rounded corners, align=center,
            minimum width=5.6em, minimum height=1.9em},
        hier/.style={thick}, mesh/.style={thick, dashed}, cflink/.style={thick, dashed, ->},
        tlbl/.style={rotate=90, font=\footnotesize\bfseries},
        band/.style={rounded corners, fill=gray!8, draw=gray!35},
      ]

      \draw[band] (-4.7,-3.125) rectangle (4.7, 1.125);
      \draw[dashed, gray!55, thick] (-4.7,-0.25) -- (4.7,-0.25);   %
      \draw[dashed, gray!55, thick] (-4.7,-1.75) -- (4.7,-1.75);   %

      \node[cc] (gtp) at (-2.4,2.15) {Group Token\\Provider};
      \node[cc] (fr)  at ( 2.4,2.15) {Function\\Registry};
      \draw[->, thick] (gtp.east) -- node[above, font=\tiny] {auth. function fetch} (fr.west);

      \foreach \name/\x/\y in {%
          cl/0/0.5, fg1/-2.6/-1.0, fg2/2.6/-1.0,
          e1/-3.9/-2.5, e2/-1.3/-2.5, e3/1.3/-2.5, e4/3.9/-2.5}{
          \node[rt] (\name) at (\x,\y) {};
          \node[font=\tiny, anchor=north] at ($(\name.north)+(0,-0.04)$) {runtime};
        }

      \node[cA] at ($(cl)+(-0.45,-0.12)$) {c1}; \node[cB] at ($(cl)+(0,-0.12)$) {c2};
      \node[cC] at ($(cl)+(0.45,-0.12)$) {c3};
      \node[cA] at ($(fg1)+(0,-0.12)$) {c1};
      \node[cC] at ($(fg2)+(0,-0.12)$) {c3};
      \node[cA] at ($(e1)+(0,-0.12)$) {c1};
      \node[cC] at ($(e3)+(0,-0.12)$) {c3};
      \node[cA] at ($(e4)+(-0.35,-0.12)$) {c1}; \node[cC] at ($(e4)+(0.35,-0.12)$) {c3};

      \draw[hier] (cl) -- (fg1);   \draw[hier] (cl) -- (fg2);
      \draw[hier] (fg1) -- (e1);   \draw[hier] (fg1) -- (e2);
      \draw[hier] (fg2) -- (e3);   \draw[hier] (fg2) -- (e4);

      \draw[cflink] (-2.4,1.125) -- node[left,  font=\tiny] {user permissions} (gtp.south);
      \draw[cflink] ( 2.4,1.125) -- node[right, font=\tiny] {function fetch}   (fr.south);

      \node[tlbl] (tl3) at (-5.5, 0.5)  {Cloud};
      \node[tlbl] (tl2) at (-5.5,-1.0)  {Fog};
      \node[tlbl] (tl1) at (-5.5,-2.5)  {Edge};

      \node[draw, very thick, rounded corners,
        fit={(gtp) (fr) (-4.7,1.125) (4.7,-3.125) (tl3) (tl1)}, inner sep=0.4em,
        label={[font=\bfseries]above:Ermes}] (ermes) {};

      \node[actor] (admin) at (-2.8,4.7) {Administrator};
      \node[actor] (dev)   at ( 2.8,4.7) {Developer};
      \draw[->, thick] (admin) -- node[left, font=\scriptsize, align=right, pos=0.55]
      {infrastructure\\deploy} (admin |- ermes.north);
      \draw[->, thick] (dev)   -- node[right, font=\scriptsize, align=left, pos=0.55]
      {function\\deploy} (dev |- ermes.north);

      \node[cli] (k1a) at (-4.3,-4.9) {client}; \node[cli] (k1b) at (-3.1,-4.9) {client};
      \node[cli] (k2)  at (-1.3,-4.9) {client};
      \node[cli] (k4a) at ( 3.1,-4.9) {client}; \node[cli] (k4b) at ( 4.3,-4.9) {client};
      \foreach \k/\e in {k1a/e1, k1b/e1, k2/e2, k4a/e4, k4b/e4}{
          \draw[->, thick] (\k) -- (\e);
        }
      \node[right, font=\tiny] at ($(k2)!0.5!(e2)$) {function invocation};

    \end{tikzpicture}}
  \caption{Functional model of the Ermes framework.}
  \label{fig:ermes-overview}
\end{figure}

Ermes aims to minimize the latency observed by clients that invoke stateful
functions from the edge of the network. This latency is dominated by two
costs: executing a function's code and accessing the state it reads and
writes. To minimize these two costs, Ermes co-locates each function with the
state it accesses both close to the client. Where a function executes and
where its state resides are thus decided jointly.
In previous work, we formulated this joint problem of function scheduling and
data placement exactly and showed that, although it admits an optimal
solution, computing one is not feasible for the edge-to-cloud
continuum~\cite{ermesTheory}; we thus devised a heuristic that approximates
the optimum well and adapts to dynamic scenarios. Ermes relies on this
heuristic.

Ermes is a distributed system: it runs on many \emph{nodes} that cooperate,
communicating and synchronizing to provide the service. The administrator
deploys the same software, the \emph{Ermes runtime}, on every node, and the
runtimes together execute the application's functions and manage the state
those functions access.
The nodes are organized into a hierarchy of \emph{tiers}, illustrated in
\fig{ermes-overview}. Computational and storage capacity grows from the leaves
toward the root, so the tiers closest to the clients are the most
resource-constrained and those higher up the most capable. A node does not
reach every other node directly: each tier connects to the one above it, and a
node can always climb toward the root to find more compute and storage. We
model the infrastructure as a tree, consistent with common edge-to-cloud
hierarchies~\cite{bonomi-fog, shi-promise}: although a physical deployment may
wire nodes differently, this logical hierarchy captures that a node reaches
richer resources by climbing toward the root. The nodes form a strict
parent-child hierarchy rooted at a single cloud node. Throughout the paper we
use a three-tier instance of this model, with an \emph{edge}, a \emph{fog},
and a \emph{cloud} tier. The collections that make up the application state
reside on these nodes, and clients issue their invocations to the edge tier
closest to them.

\begin{figure}[tbp]
  \centering
  \resizebox{\columnwidth}{!}{%
    \begin{tikzpicture}[>=Stealth, font=\small,
        nd/.style={draw, thick, rounded corners=1pt, fill=white,
            minimum width=1.05cm, minimum height=0.66cm, font=\footnotesize},
        coll/.style={draw, thick, cylinder, shape border rotate=90,
            aspect=0.28, minimum width=1.0em, minimum height=0.9em,
            inner sep=1pt, font=\tiny},
        cA/.style={coll, fill=blue!18}, cB/.style={coll, fill=green!22},
        cC/.style={coll, fill=orange!28},
        lnk/.style={-, thick},
        lbl/.style={font=\footnotesize},
        ldr/.style={star, star points=5, star point ratio=2.25, fill=black, inner sep=0.8pt}]

      \draw[dashed, rounded corners=2pt] (-0.9, 0.5) rectangle (0.9, 2.45);
      \node[lbl] at (0, 2.65) {parent};

      \node[nd] (p2) at (0,2) {};
      \node[nd] (p) at (0,1) {};
      \node[nd, fill=blue!8, label={[lbl]below:node $n$}] (n) at (0,0) {};
      \node[nd, label={[lbl]below:child$_1$}] (c1) at (-1.7,-0.8) {};
      \node[nd, label={[lbl]below:child$_2$}] (c2) at (1.7,-0.8) {};

      \draw[lnk] (n) -- (p);
      \draw[lnk] (p) -- (p2);
      \draw[lnk] (n) -- (c1);
      \draw[lnk] (n) -- (c2);

      \node[cA] (c1_p2) at (p2.center) {$c_1$};
      \node[ldr] at ($(c1_p2.east)+(0.15,0)$) {};
      \node[cC] (c3_p) at (p.center) {$c_3$};
      \node[ldr] at ($(c3_p.east)+(0.15,0)$) {};
      \node[cB] at (n.center) {$c_2$};
      \node[cA] at (c1.center) {$c_1$};
      \node[cB] at (c2.center) {$c_2$};

      \node[anchor=west, align=left, font=\footnotesize, draw, rounded corners=2pt, fill=blue!8]
      (tbl) at (2.5, 0) {%
        \begin{tabular}{@{}lll@{}}
          \textit{collection} & \textit{replica direction} & \textit{leader direction} \\ \hline
          $c_1$               & parent, child$_1$          & parent                    \\
          $c_2$               & local, child$_2$           & ---                       \\
          $c_3$               & parent                     & parent                    \\
        \end{tabular}};

      \draw[->, dashed, thick, draw=black!70] (n.east) -- (tbl.west);

    \end{tikzpicture}}
  \caption{Node $n$ and the reachability information it keeps; a $\bigstar$
    marks a leader.}
  \label{fig:reachability}
\end{figure}
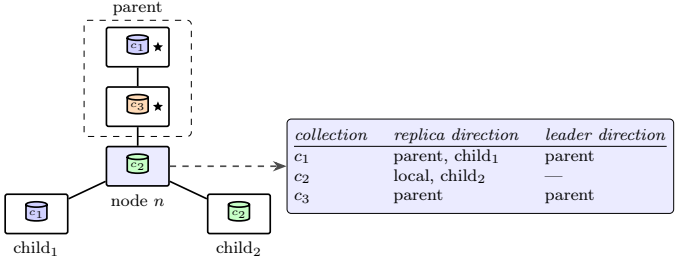

Two centralized components complete the deployment.
The \emph{function registry} stores all the functions available in the
application, so that any node can retrieve a function it does not yet hold
locally.
The \emph{group token provider} is the sole authority on the group memberships
of clients: upon authentication, it issues a signed \emph{group token} with a
limited lifetime, which the client attaches to its invocation requests, and
which any node can verify locally to enforce the access policies of the target
collections without contacting the provider on every request.

No node holds a global view of the system. Instead, each node maintains
\emph{reachability information} (\fig{reachability}), used to provide its
functionalities: for every collection it is aware of, the direction in which a
replica lies, toward its parent or into a specific child's subtree, and, for a
sequentially consistent collection, the direction toward its leader. Note that
direction and location differ: in \fig{reachability}, node $n$ knows $c_1$
lies toward its parent, though its replica actually sits two tiers up. A node
assembles this view by interacting with its neighbors.
We assume the topmost tier, the cloud, retains a replica of every collection:
with effectively unbounded storage it can hold them all, so a node always
reaches a collection by climbing toward the root. This makes the cloud the
fallback from which every collection is reachable.

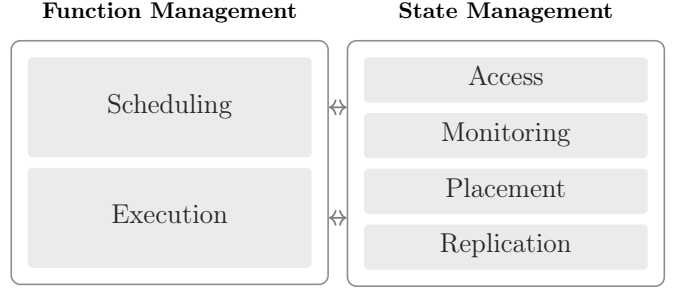
\begin{figure}[tbp]
  \centering
  \resizebox{\columnwidth}{!}{
    \begin{tikzpicture}[ font=\sffamily, gap/.store in=\gap, gap=2.5mm,
        layer/.style={fill=black!8, rounded corners=2pt, text=black!85,
            minimum width=4.4cm, font=\large, text height=1.6ex, text depth=0.5ex, align=center},
        layerSM/.style={layer, minimum height=0.7cm},
        layerFM/.style={layer, minimum height=1.55cm},
        box/.style={draw=black!45, line width=0.7pt, rounded corners=4pt, inner sep=\gap} ]

      \node[layerSM] (dist) {Access};
      \node[layerSM, below=1.5mm of dist] (mon) {Monitoring};
      \node[layerSM, below=1.5mm of mon] (place) {Placement};
      \node[layerSM, below=1.5mm of place] (repl) {Replication};

      \node[layerFM, left=8mm of dist.north west, anchor=north east] (sched) {Scheduling};
      \node[layerFM, below=1.5mm of sched] (exec) {Execution};

      \node[box, fit=(sched) (exec)] (fe_box) {};
      \node[box, fit=(dist) (repl)] (sm_box) {};

      \node[font=\bfseries, above=1.5mm of fe_box] {Function Management};
      \node[font=\bfseries, above=1.5mm of sm_box] {State Management};

      \draw[<->, thick, draw=black!50]
      (fe_box.east |- sched.center) -- (sm_box.west |- sched.center);
      \draw[<->, thick, draw=black!50]
      (fe_box.east |- exec.center) -- (sm_box.west |- exec.center);

    \end{tikzpicture}
  }
  \caption{The functionalities of Ermes.}
  \label{fig:ermes-func}
\end{figure}

\fig{ermes-func} organizes the runtime's functionalities into two groups that
cooperate, each realized in a distributed fashion through protocols among the
runtimes, with no central coordinator. \emph{Function management} brings each
invocation to a node and runs it there, through the \emph{scheduling} and
\emph{execution} functionalities. \emph{State management} keeps the
collections spread across the hierarchy and serves the accesses functions make
on them, through the \emph{access}, \emph{monitoring}, \emph{placement}, and
\emph{replication} functionalities.

When a client invokes a function, it contacts a node of the hierarchy,
typically the closest one, which it selects itself. From that node,
\emph{scheduling} decides where the invocation is served, locally or forwarded
to another node, and \emph{execution} then runs the function there.

Functions are stateful and operate on collections. A collection exists as a
set of replicas whose roles follow its consistency policy: a sequentially
consistent collection has a single leader replica that accepts writes and any
number of read-only followers, while an eventually consistent one is
multi-leader, every replica accepting both reads and writes. While a function
runs, the serving node resolves the collections it needs and carries out its
reads and writes (\emph{access}). In the background, \emph{monitoring} records
how much demand each collection draws and from which direction, and
\emph{placement} consumes this to continuously reconsider where each
collection and its replicas should reside, migrating them as the demand of the
clients evolves. Finally, \emph{replication} keeps the replicas of a
collection synchronized, both while functions update them and while they
migrate.

The two families are the subject of the two sections that follow: function
management in \s{func:mng} and state management in \s{state:mng}. %
\section{Function Management}
\label{sec:func:mng}
Function management covers the path of an invocation, from the moment a client
issues it to the moment the platform returns its response. It brings each
invocation to a node and runs the function there, through the
\emph{scheduling} and \emph{execution} functionalities. Its half of the
co-location goal (\s{sys_overview}) is to serve each invocation on a node that
already holds the data it needs, so that the function runs beside its state
and the client observes low latency. Because the decision sits on the critical
path of every invocation, a node never consults nodes beyond its neighbors,
and either executes the invocation or forwards it one hop toward the data. And
because collections migrate across the hierarchy while invocations are in
flight (\s{state:mng}), it decides against a placement that is never fixed.

\fakeparagraph{Scheduling}
The node a client contacts is not necessarily the one that should execute the
invocation, as the data it needs may be hosted elsewhere. To route without
coordination, each node relies on the \emph{reachability information} it
maintains locally (\s{sys_overview}, \fig{reachability}): for every collection
it is aware of, the direction in which a replica is hosted and, for a
sequentially consistent collection, the direction toward its leader. This view
is partial, as a node is aware only of the collections within its reach; a
node that knows nothing of a required collection forwards the invocation to
its parent, which is aware of more, up to the root, which is aware of every
collection.

When an invocation enters the system, its entry node resolves the set $C$ of
collections it requires (\s{prog_model}). If $C$ is empty, nothing constrains
the execution and the node serves the invocation itself. Otherwise, $n$ looks
each collection up in its reachability information: for a collection $c$ it
reads the neighbors $\mathrm{repl}(c)$ toward which a replica is hosted
($\textsf{self}$ if $n$ holds one) and, for a sequentially consistent $c$, the
neighbor $\mathrm{lead}(c)$ toward its leader; it also knows the latency
$\mathit{lat}(e)$ to each neighbor $e$, its parent and children
($\mathit{lat}(\textsf{self}) = 0$), and treats a collection it is unaware of
as reachable through its parent. From these it decides as \algo{scheduling}
summarizes, following the consistency policy of the collections involved.

\begin{algorithm}
  \caption{Scheduling decision at a node $n$, computing the direction $d$ it
    serves the invocation toward.}
  \label{algo:scheduling}
  \begin{algorithmic}[1]
    \REQUIRE the required collections $C$
    \IF{$C = \emptyset$}
    \STATE $d \gets \textsf{self}$
    \ELSIF{some $c \in C$ is sequentially consistent}
    \IF{the function writes $c$}
    \STATE $d \gets \mathrm{lead}(c)$
    \ELSE
    \STATE $d \gets \arg\min_{e \,\in\, \mathrm{repl}(c)} \mathit{lat}(e)$
    \ENDIF
    \ELSIF{$\textsf{self} \in \mathrm{repl}(c)$ for some $c \in C$}
    \STATE $d \gets \textsf{self}$
    \ELSE
    \STATE $d \gets \arg\max_{e \,\in\, \bigcup_{c \in C} \mathrm{repl}(c)}
      \lvert \{\, c \in C : e \in \mathrm{repl}(c) \,\} \rvert$
    \ENDIF
    \IF{$d = \textsf{self}$}
    \RETURN \textsc{execute}
    \ELSE
    \RETURN \textsc{forward} toward $d$
    \ENDIF
  \end{algorithmic}
\end{algorithm}
A sequentially consistent collection dominates the decision, since it alone
constrains where execution may run: a write must reach the single leader,
while a read may go to any replica, so $n$ heads toward the nearest one. With
only eventually consistent collections, $n$ instead maximizes co-location,
executing locally when it already holds a required collection and otherwise
moving toward the neighbor that covers most of them. Any eventually consistent
collection still missing at the chosen node is fetched from the nearest known
replica while the function runs.

Because each node either executes or forwards one hop, an invocation climbs at
most to the root, so routing always terminates. How often it must climb far
depends on where replicas sit, which is exactly what collection placement
(\s{state:mng}) works to improve, by keeping data close to the clients that
use it.

\fakeparagraph{Execution}
Once scheduling has selected the node that serves an invocation, that node
executes the requested function locally. Two requirements shape how it does
so.

First, a node serves concurrently the invocations of different functions, on
behalf of different clients: each invocation must therefore run in isolation
from the others.
Second, scheduling may direct an invocation to any node of the continuum,
whose hardware ranges from constrained edge devices to cloud servers: the same
function must thus be able to run, unmodified, wherever it lands.
Ermes accordingly runs every invocation in an isolated execution environment,
uniform across the infrastructure and private to the invocation it serves. It
realizes this environment on \emph{WASP}~\cite{wasp}, a configurable substrate
for stateful serverless execution in the edge-to-cloud continuum, whose
implementation we detail in \s{implementation}.
\section{State Management}
\label{sec:state:mng}
Function management brings each invocation to a node and runs it there; it
does so over a body of state that state management maintains. Ermes holds the
application state as collections and continually replicates and relocates them
across the hierarchy, so that each collection sits close to the clients that
use it. State management is in this sense the counterpart of function
management, and the two reinforce each other: placement pulls a collection
toward the nodes that request it, shortening the path invocations travel,
while the traffic that scheduling routes drives placement.
State management comprises four functionalities (\fig{ermes-func}):
\begin{inparaenum}[(i)]
  \item \emph{access} serves the reads and writes that functions perform on
  collections,
  \item \emph{monitoring} measures the demand each collection draws,
  \item \emph{placement} decides where each collection and its replicas reside
  as that demand shifts, and
  \item \emph{replication} keeps the replicas of a collection consistent as
  functions update them and as they move.
\end{inparaenum}
Access runs on the critical path of every invocation; the other three run in
the background.

\fakeparagraph{Access}
Ermes enforces the access policy of each collection (\s{prog_model}) with a
purely local check. The client proves its group memberships through the group
token issued by the group token provider (\s{sys_overview}), and the node
serving an access verifies that token against the groups authorized on the
collection, deciding locally and with no lookup elsewhere. Both the authorized
groups of a collection and the memberships of a client may change over time,
and each access is decided against their state at the moment it is served.

The authorized groups are part of the \emph{collection metadata}
(\s{prog_model}): when an invocation enters the system, the platform resolves
the collections the function declares against this metadata (\s{func:mng}).

Once the collections are resolved, the function reads and writes their records
through the \texttt{get} and \texttt{put} operations, unaware of where the
collections physically reside. When a required replica is on the local node,
the operation is served there; when it is not, the interface transparently
forwards the operation to a node that holds one.  How these operations leave
the replicas of a collection mutually consistent is the subject of the
replication functionality below.

\fakeparagraph{Monitoring}
Placement needs to know how much demand each collection draws and from
where;monitoring supplies it, continuously and off the critical path. Whenever
a node finishes executing a function, it emits an \emph{access report} for
each collection the function accessed, recording whether the access was a read
or a write and the latency the invocation accumulated on its way, one link at
a time. Each report is routed toward a node that holds the collection,
forwarded hop by hop by the same reachability information that guides
scheduling (\s{func:mng}). As the reports travel through the hierarchy, each
node keeps, for every collection it hosts, two running sums over the reports
it receives: their number and the total latency they carry.

\fakeparagraph{Placement}
A collection is the atomic unit of storage and replication: its records are
never partitioned across nodes but reside together, so that a replica is
self-contained and placement moves or replicates it as a whole. Each node
periodically decides the placement of the collections it holds, weighing the
demand monitoring has gathered against the free storage of its neighbors.
These local decisions realize the decentralized heuristic that our previous
work~\cite{ermesTheory} derives for the placement problem (\s{sys_overview});
here we describe how that heuristic is mapped onto Ermes.

\begin{figure}[tbp]
  \centering
  \begin{tikzpicture}[>=Stealth,
      phase/.style={draw, rounded corners, minimum width=5em,
          minimum height=2.2em, align=center, font=\small},
      tlbl/.style={font=\scriptsize, align=center},
      slbl/.style={font=\scriptsize\itshape, align=center, text=black!65}]
    \node[phase] (idle) at (0,0) {Idle};
    \node[phase] (gossip) at (3.2,0) {Gossip};
    \node[phase] (decision) at (6.4,0) {Decision};
    \draw[->] (idle) -- node[above=4.5mm, tlbl] {timer fires $\lor$\\capacity report} (gossip);
    \draw[->] (gossip) -- node[above=4.5mm, tlbl] {capacity reports\\from all neighbors} (decision);
    \draw[->] (decision.south) to [bend left=25]
    node[below, tlbl] {decisions taken} (idle.south);
  \end{tikzpicture}
  \caption{The placement state machine run by each node.}
  \label{fig:placement-fsm}
\end{figure}
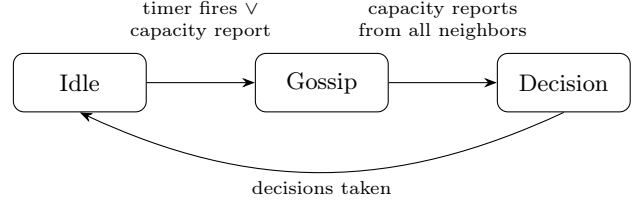

A node acts on the accumulated demand periodically, through a cyclic state
machine with three phases, \emph{idle}, \emph{gossip}, and \emph{decision}; we
call one full cycle an \emph{epoch} (\fig{placement-fsm}).

Initially, the node remains \emph{idle} for a fixed time, where it just serves
invocations. This phase bounds how often the machine runs, so that epochs do
not follow one another too closely. Two conditions drive the transition to the
\emph{gossip phase}: either a timeout fires after the fixed time, or a
\emph{capacity report} from a neighbor arrives. In this way, epochs stay
loosely aligned across the tree with no global clock.

Entering the \emph{gossip} phase, the node sends its residual storage capacity
to both its parent and children, and waits to receive theirs.

Once it holds the capacity report of every neighbor, the node enters the
\emph{decision} phase. It evaluates each collection it holds independently,
weighing, for every neighbor, the free storage it has just learned against the
demand it has accumulated, and decides whether to keep the collection in
place, migrate it to a neighbor, or replicate it. The node then returns to
idle, and the migrations originated in the last phase proceed asynchronously.

The two consistency policies supported by Ermes, call for different placement
mechanisms, because they differ in whether the placement choices for a
collection are coupled.

Under eventual consistency placement is \emph{reactive}: any replica accepts
writes, so each one can be placed independently. Placement piggybacks on the
natural flow of requests: when a function accesses a collection not stored
locally, its node fetches it from the nearest known replica, and, as part of
serving that request, the holder asynchronously provisions a new copy one hop
closer to the consumer. Over time, collections under sustained demand migrate
hop by hop toward the edge nodes that access them most. The decision to copy
is subject to a storage threshold: a node sends a copy to a neighbor only if
the neighbor has sufficient free capacity.

Under sequential consistency, instead, placement is \emph{proactive}: every
write must reach the single leader, dictating where write traffic flows and
where followers are worth placing. Ideally, the leader should be as close as
possible to the clients issuing writes; but edge nodes have limited storage,
and accumulating all replicas near the leaves is infeasible. Placement
therefore balances bringing state close to clients against the storage
constraints of resource-scarce nodes.

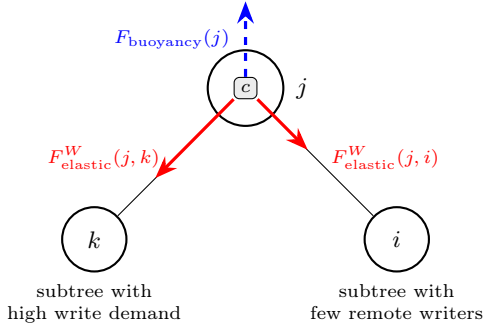
\begin{figure}[tbp]
  \centering
  \begin{tikzpicture}[>=Stealth, font=\small, host/.style={circle, draw,
          thick, fill=white}, rep/.style={draw, rounded corners=2pt, inner
          sep=2.5pt, font=\scriptsize\sffamily, fill=gray!15},
      lbl/.style={font=\scriptsize, align=center}]

    \node[host, minimum size=1cm] (nj) at (2.6, 0.2) {}; \node[rep] (rep) at
    (nj) {$c$}; \node[font=\small, right=1pt of nj] {$j$}; \node[host, minimum
      size=0.85cm] (c1) at (0.6, -1.8) {$k$}; \node[host, minimum size=0.85cm]
    (c2) at (4.6, -1.8) {$i$}; \node[lbl, below=1pt of c1] {subtree with\\high
      write demand}; \node[lbl, below=1pt of c2] {subtree with\\few remote
      writers}; \draw (nj) -- (c1); \draw (nj) -- (c2);

    \draw[->, very thick, shorten >=20pt, red] (rep.south west) -- (c1.north
    east) node[midway, left=2pt, lbl] {$F^W_{\text{elastic}}(j,k)$};
    \draw[->, very thick, shorten >=35pt, red] (rep.south east) --
    (c2.north west) node[midway, right=2pt, lbl]
    {$F^W_{\text{elastic}}(j,i)$}; \draw[->, very thick, dashed,
      blue] (rep.north) -- ++(0, 1) node[midway, left=2pt, lbl]
    {$F_{\text{buoyancy}}(j)$};
  \end{tikzpicture}
  \caption{The leader replica of collection $c$ on node $j$ is pulled toward
    child $k$ more than toward sibling $i$, which write less, and pushed upward
    by the buoyant force of a nearly-full node.}
  \label{fig:placement-forces}
\end{figure}

The heuristic casts the balance between locality and storage as a system of
virtual forces on each replica (\fig{placement-forces}). Demand acts like a
spring: it draws the replica toward the clients it serves, more strongly the
more requests they issue and the farther away they are.
The elastic force of a direction is exactly the total latency the node has
accumulated for it. For a neighbor $k$, \eqref{eq:forces} factors this total
into the number of requests $\lambda_k$ that reach the collection from $k$'s
subtree and the latency $d_{jk}$ of the link between the node $j$ and $k$.
Recall that writes are directed only to the leader while reads can be served
by any replica, so this pull splits in two: write demand anchors the leader,
read demand decides where followers are placed. We accordingly split the count
$\lambda_k$ into the write and read counts $\lambda^W_k$ and $\lambda^R_k$
aggregated from neighbor $k$, and write $F^W_{\text{elastic}}(j,k)$ and
$F^R_{\text{elastic}}(j,k)$ for the elastic force each induces.
The opposing force is storage: a node $j$ with little free storage behaves
like a dense fluid that pushes its replicas up toward the parent, where free
storage is more abundant. Its magnitude $F_{\text{buoyancy}}(j)$ grows with
the size $\mathit{size}_c$ of the collection $c$ and with the occupancy of the
node, as the reciprocal of its free storage $\mathit{stor}^{\text{free}}_j$,
which a node learns through gossip. The constant $\mu$ places this force on
the same scale as the elastic one.

  {\footnotesize
    \begin{align}
      F_{\text{elastic}}(j,k) = \lambda_k \cdot d_{jk},
      \qquad
      F_{\text{buoyancy}}(j) = \frac{\mu}{\mathit{stor}^{\text{free}}_j}
      \cdot \mathit{size}_c
      \label{eq:forces}
    \end{align}}

At equilibrium, the leader is positioned to minimize write latency and the
followers to minimize read latency, and the two positions are determined by
separate conditions.
The leader migrates one hop at a time, and each node re-evaluates the handoff
to every child once per epoch. Consider the leader at node $j$ in
\fig{placement-forces}, where a write-heavy child $k$ pulls it downward while
the lighter siblings resist the move: relocating the leader to $k$ reduces
latency for the writes originating in $k$'s subtree, but increases it for all
remaining writers, namely the siblings and the demand reaching $j$ from above.
Node $j$ does not track this external demand explicitly: since every write
reaches the leader, the demand originating outside $j$'s subtree arrives
precisely through its parent link, and the force $j$ measures on that link
summarizes it in a single scalar, the \emph{tension} $F_{\text{ext},j}^W$.
Node $j$ delegates the leader to $k$ when the inward pull of $k$'s subtree,
net of the sibling pulls and the tension, exceeds the cost of placing the
replica one tier lower, namely its buoyancy at $k$ and the extra link $d_{jk}$
that each of the $\lambda^W_i$ writes of the siblings would now traverse, by a
hysteresis margin $\tau_L$ that prevents oscillation between adjacent
nodes~\eqref{eq:leader-delegation}.

  {\footnotesize
    \begin{align}
       & F^W_{\text{elastic}}(j,k)
      - \!\!\sum_{\substack{i \in \mathrm{ch}(j) \\ i \neq k}}\!\!
      F^W_{\text{elastic}}(j,i) - F_{\text{ext},j}^W
      \nonumber                          \\
       & \qquad > F_{\text{buoyancy}}(k)
      + \!\!\sum_{\substack{i \in \mathrm{ch}(j) \\ i \neq k}}\!\!
      \lambda^W_i \cdot d_{jk} + \tau_L
      \label{eq:leader-delegation}
    \end{align}}

The delegation recurses with no additional bookkeeping. Once the leader
resides on $k$, the writes of the rest of the tree keep flowing to it, now
traversing the link between $j$ and $k$: the tension $k$ measures on its
parent link therefore amounts to~\eqref{eq:ancestor-tension}, the sibling
pulls left at $j$, their extra hop, and the tension $j$ was itself subject to.
Node $k$ can thus apply the same test to its own children, and the leader
descends toward the region of highest write demand as long as the condition
holds.

  {\footnotesize
    \begin{align}
      F_{\text{ext},k}^W \gets
      \!\!\sum_{\substack{i \in \mathrm{ch}(j) \\ i \neq k}}\!\!
      F^W_{\text{elastic}}(j,i)
      + \!\!\sum_{\substack{i \in \mathrm{ch}(j) \\ i \neq k}}\!\!
      \lambda^W_i \cdot d_{jk}
      + F_{\text{ext},j}^W
      \label{eq:ancestor-tension}
    \end{align}}

Follower placement requires no such coordination. Because any replica can
serve reads, a follower benefits only the subtree beneath it, so each node
evaluates each child independently, without sibling or tension terms. Node $j$
provisions a follower on child $k$ when the read demand $\lambda^R_k$
aggregated from $k$'s subtree exceeds the storage pressure at $k$ by a margin
$\tau_R$~\eqref{eq:follower-replication}.

  {\footnotesize
    \begin{align}
      F^R_{\text{elastic}}(j,k) > F_{\text{buoyancy}}(k)
      + \tau_R
      \label{eq:follower-replication}
    \end{align}
  }

Both mechanisms replicate aggressively toward the edge, but storage is finite,
so placement is complemented by a hierarchical eviction that reclaims it.
Each node keeps its replicas in least-recently-used (LRU) order and evicts
them, on two occasions: reactively, when an incoming collection does not fit
in the residual space, the node evicts replicas until it does; and
proactively, at the end of each decision phase, when occupancy exceeds a
configured safe level, the node migrates enough replicas to its parent to fall
back below it.
Evicted replicas are not discarded but pushed to the parent. A node that
receives an evicted replica absorbs it if it has spare capacity, and otherwise
pushes it further up, until a node with room takes it in, or it reaches the
cloud, whose storage is unbounded. A node that receives a replica it already
holds simply discards the incoming copy.

\fakeparagraph{Replication}
Ermes keeps replicas synchronized, as functions update them and as placement
provisions and moves them.

Read operations are served synchronously during the execution. Writes,
instead, never reach the store while the function runs: they are accumulated
in a per-collection set of dirty records, which the function hands back to the
node together with its response when it returns. Only then the node commits
atomically each set, as a single batch per collection, so that a failed
execution leaves no partial writes behind.

The two policies commit at different moments. The batch of the sequentially
consistent collection is committed before the response returns to the client.
The batch of the eventually consistent collections is instead committed
asynchronously, after the response has returned, so the client never waits for
them.

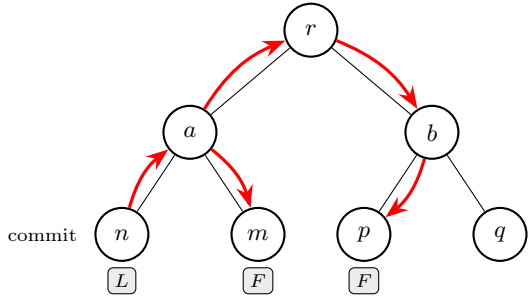
\begin{figure}[tbp]
  \centering
  \begin{tikzpicture}[>=Stealth, font=\small, host/.style={circle, draw,
          thick, fill=white, minimum size=0.7cm, inner sep=1pt},
      rep/.style={draw, rounded corners=2pt, inner sep=2.5pt,
          font=\scriptsize\sffamily, fill=gray!15}, upd/.style={->, very
          thick, red}, lbl/.style={font=\scriptsize, align=center}]
    \node[host] (r) at (2.7, 2.7) {$r$}; \node[host] (a) at (1.1, 1.35)
    {$a$}; \node[host] (b) at (4.3, 1.35) {$b$}; \node[host] (n) at
    (0.2, 0) {$n$}; \node[host] (m) at (2.0, 0) {$m$}; \node[host] (p)
    at (3.4, 0) {$p$}; \node[host] (q) at (5.2, 0) {$q$}; \draw (r) --
    (a); \draw (r) -- (b); \draw (a) -- (n); \draw (a) -- (m); \draw (b)
    -- (p); \draw (b) -- (q); \node[rep, below=2pt of n] {$L$};
    \node[rep, below=2pt of m] {$F$}; \node[rep, below=2pt of p] {$F$};
    \node[lbl, left=3pt of n] {commit}; \draw[upd] (n) to[bend left=18]
    (a); \draw[upd] (a) to[bend left=18] (r); \draw[upd] (a) to[bend
      left=18] (m); \draw[upd] (r) to[bend left=18] (b); \draw[upd] (b)
    to[bend left=18] (p);
  \end{tikzpicture}
  \caption{Up-and-down propagation of an update committed at the leader
    replica ($L$) on node $n$. Nodes $a$, $r$, and $b$ hold no replica and
    forward without applying; the subtree of $q$ holds no replica and is never
    reached.}
  \label{fig:sync-up-down}
\end{figure}

Once committed on the executing node, an update disseminates through an
\emph{up-and-down} protocol rooted at that node (\fig{sync-up-down}). Each
node forwards the update in every direction its reachability information
(\s{sys_overview}) indicates a replica, excluding only the link the update
arrived from, and applies it locally if it holds a replica itself. The updates
directed to the same neighbor are delivered in order, one at a time.

For sequentially consistent collections, at each commit, the leader increments
a per-collection counter and tags the outgoing update with the resulting
\emph{sequence number}. Because a node acknowledges an update only after
applying and forwarding it, and per-neighbor delivery is ordered, updates
reach every follower in sequence-number order; a replica moreover applies an
incoming update only if its sequence number exceeds the local one, discarding
the duplicate and stale deliveries that can arise while replicas move, so
every replica evolves monotonically toward the leader state.
Reads can be served by any replica. Each response reports the sequence number
the serving replica had reached, and each request carries the highest number
the client has observed so far: a replica behind that number does not serve
the invocation immediately, but re-examines its local state with an
exponential backoff until it catches up, aborting the invocation after a
bounded number of attempts. This preserves read-your-writes even when
consecutive invocations of a moving client are served by different replicas.

For eventually consistent collections, any replica accepts reads and writes,
with no coordination on the critical path. Concurrent updates to the same
record are resolved by a per-record \emph{last-writer-wins} (LWW) rule, where
each update is timestamped at its origin, and ties are broken with the
identifier of the originating node. The comparison and the write execute as a
single atomic action on the local store, so that concurrent appliers cannot
interleave. The rule totally orders the updates of each record, and replicas
that have observed the same set of updates hold the same state.

\begin{figure}[tbp]
  \centering
  \subfloat[An update racing the transfer on the same
    link.\label{fig:sync-race-same}]{%
    \resizebox{0.47\linewidth}{!}{%
      \begin{tikzpicture}[>=Stealth, font=\scriptsize, msg/.style={->, thick},
          life/.style={gray!60}, evt/.style={draw, rounded corners=2pt, inner
              sep=2.5pt, font=\scriptsize\sffamily, fill=gray!15}]

        \useasboundingbox (-1.1, 0.2) rectangle (3.5, -4.2);

        \node[font=\small] at (0,0) {$s$}; \node[font=\small] at (2.6,0)
        {$d$}; \draw[life] (0,-0.25) -- (0,-4.1); \draw[life] (2.6,-0.25) --
        (2.6,-4.1); \draw[msg] (0,-0.5) -- node[above, sloped] {transfer $c$}
        (2.6,-1.7); \node[evt, anchor=east] at (-0.08,-1.2) {postpone $u$};
        \node[evt, anchor=west] at (2.68,-1.85) {store $c$}; \draw[msg]
        (2.6,-2.1) -- node[above, sloped] {ack} (0,-2.75); \draw[msg] (0,-3.0)
        -- node[above, sloped] {$u$} (2.6,-3.6); \node[evt, anchor=west] at
        (2.68,-3.85) {apply $u$}; \end{tikzpicture}%
    }%
  }
  \hfill
  \subfloat[An update and a replica crossing.\label{fig:sync-race-cross}]{%
    \resizebox{0.47\linewidth}{!}{%
      \begin{tikzpicture}[>=Stealth, font=\scriptsize, msg/.style={->, thick},
          life/.style={gray!60}, evt/.style={draw, rounded corners=2pt, inner
              sep=2.5pt, font=\scriptsize\sffamily, fill=gray!15}]

        \useasboundingbox (-1.1, 0.2) rectangle (3.5, -4.2);

        \node[font=\small] at (0,0) {$x$}; \node[font=\small] at (2.6,0)
        {$y$}; \draw[life] (0,-0.25) -- (0,-4.1); \draw[life] (2.6,-0.25) --
        (2.6,-4.1); \draw[msg] (0,-0.5) -- node[above, sloped, pos=0.28] {$u$}
        (2.6,-2.3); \draw[msg] (2.6,-0.9) -- node[above, sloped, pos=0.28]
        {transfer $c$} (0,-2.0); \node[evt, anchor=east] at (-0.08,-2.25)
        {store $c$}; \draw[msg] (2.6,-2.55) -- node[above, sloped] {ack}
        (0,-3.3); \node[evt, anchor=east] at (-0.08,-3.6) {re-apply $u$};
      \end{tikzpicture}%
    }%
  } \caption{The two races between an update $u$ for a collection $c$ and a
    transfer of $c$. Arrows are messages exchanged between the two nodes;
    boxes are local actions on a node's timeline.}
  \label{fig:sync-races}
\end{figure}
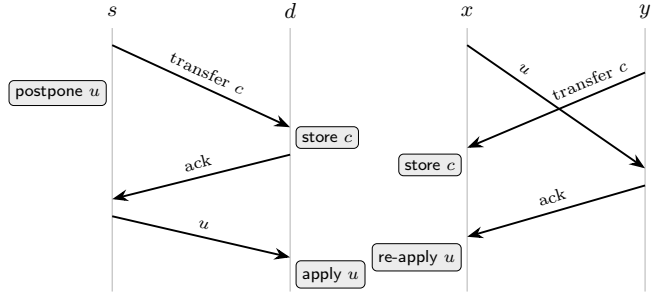

Placement moves replicas while updates are in flight, and a transfer and an
update may travel the same link at the same time, in the same direction or in
opposite ones; without precautions, both races could lose updates permanently.
The protocol prevents this with two dual safeguards.
A node that is transferring a replica postpones every update directed to the
transfer destination: instead of shipping it, the node enqueues it, per
destination and in arrival order (\fig{sync-race-same}). If the destination
acknowledges the transfer, the node ships the postponed updates in their
original order, so they reach the new replica right after the state they were
racing; if the transfer fails, the node restores its local replica and applies
the updates to it, again in order, and marks the destination as saturated
until the next gossip round refreshes it.
Dually, a node that propagates an update for a collection it does not hold
keeps the update alive while the propagation is in flight
(\fig{sync-race-cross}): a completion counter, decremented as each outgoing
message is delivered, re-applies the update to the local store after the last
delivery if a replica of the collection has arrived in the meantime, as that
replica may have left its source before the update reached it.

A leader is not tied to a node forever: placement can migrate it, and while it
moves the collection risks being left with no leader, or with two at once.
Ermes transfers leadership through a strict handoff that rules this out. The
source demotes itself and records the destination as the new leader, and only
then starts the transfer; the destination assumes leadership upon receiving
the collection, and if the transfer fails, the source resumes it. At most one
node is thus the leader at any time.
A handoff never interrupts a running function: if placement decides one while
invocations are accessing the collection, the move waits for them to finish.
The price is a brief unavailability window, bounded by the invocations already
admitted and by the transfer itself, during which invocations reaching the
collection are rejected and succeed upon retry. Follower provisioning, in
contrast, requires no handoff: a new follower is initialized from an existing
replica and then receives subsequent updates through the regular propagation
channel, without ever blocking ongoing writes.
\section{Implementation}
\label{sec:implementation}
We implemented Ermes in Go, whose lightweight concurrency suits the
asynchronous, message-driven nature of the platform. We describe how the two
families of functionalities are realized.

\subsection{Function Management}
\label{sec:impl:func}

Every node runs an instance of the Ermes runtime, which exposes a single HTTP
endpoint: through it, the runtime receives both the invocations that clients
issue directly to the node and those that a neighboring node forwards to it as
the outcome of its scheduling decisions.

Ermes builds execution on \emph{WASP} (\s{func:mng}), which runs every function
in a WebAssembly (WASM)~\cite{webassembly} sandbox. WASM meets the two
requirements of the execution functionality: its sandbox isolates each
invocation from the others the node serves, and its portability lets the same function
binary run on any node of the continuum.

\begin{figure}[tbp]
  \centering
  \begin{tikzpicture}[
      >=Stealth,
      font=\small,
      layer/.style={draw, rounded corners=2pt, minimum width=6.3cm, minimum height=0.8cm},
      tag/.style={font=\scriptsize\bfseries, anchor=west, align=left},
      sub/.style={font=\scriptsize\itshape, anchor=east},
      chip/.style={draw, rounded corners=1pt, fill=white, font=\scriptsize\ttfamily, inner xsep=2pt, inner ysep=1.5pt},
      wbox/.style={draw, rounded corners=2pt, minimum height=1.05cm},
      flow/.style={->, thick},
      hit/.style={->, semithick},
      slbl/.style={font=\scriptsize, align=left, inner sep=2pt}
    ]

    \node[layer, dashed, draw=gray!70, fill=gray!6] (reg) {};
    \node[tag] at ([xshift=2mm]reg.west) {Function registry};
    \node[sub] at ([xshift=-2mm]reg.east) {remote};

    \node[layer, fill=gray!12, below=0.5cm of reg] (t1) {};
    \node[tag] at ([xshift=2mm]t1.west) {Tier 1\\[-2pt]\mdseries\itshape binaries, on disk};
    \node[chip] (b2) at ([xshift=-2mm]t1.east) {f@v2};
    \node[chip, left=1.2mm of b2] (b1) {f@v1};

    \node[layer, fill=gray!12, below=0.5cm of t1] (t2) {};
    \node[tag] at ([xshift=2mm]t2.west) {Tier 2\\[-2pt]\mdseries\itshape modules, in memory};
    \node[chip, fill=blue!12] (m2) at ([xshift=-2mm]t2.east) {f@v2};
    \node[chip, fill=blue!12, left=1.2mm of m2] (m1) {f@v1};

    \node[layer, fill=gray!6, minimum height=1.7cm, below=0.65cm of t2] (pool) {};
    \node[tag, anchor=north west, font=\scriptsize\bfseries] at ([xshift=1.5mm,yshift=-1mm]pool.north west) {Worker pool};

    \node[wbox, fill=green!10, minimum width=2.7cm, anchor=south west] (wa) at ([xshift=2mm,yshift=2mm]pool.south west) {};
    \node[font=\tiny, anchor=north] at ([yshift=-0.5mm]wa.north) {busy: instance of \texttt{f@v1}};
    \node[chip, dashed, fill=blue!12, anchor=north, minimum width=2.3cm, font=\tiny\ttfamily] (code) at ([yshift=-3.4mm]wa.north) {code (shared)};
    \node[chip, fill=orange!25, anchor=north, minimum width=2.3cm, font=\tiny\ttfamily] (mem) at ([yshift=-0.6mm]code.south) {linear memory (fresh)};

    \node[wbox, fill=white, minimum width=0.9cm, right=2.2mm of wa, font=\scriptsize] (i1) {idle};
    \node[wbox, fill=white, minimum width=0.9cm, right=1.8mm of i1, font=\scriptsize] (i2) {idle};
    \node[wbox, fill=white, minimum width=0.9cm, right=1.8mm of i2, font=\scriptsize] (i3) {idle};

    \draw[flow] (reg) -- node[slbl, right] {fetch} (t1);
    \draw[flow] (t1) -- node[slbl, right] {compile} (t2);
    \draw[flow] (code.north west |- t2.south) -- node[slbl, fill=white, right] {instantiate} (code.north west |- pool.north);
    \draw[flow] (wa.east |- pool.north)
    -- ($(wa.east |- pool.north)+(0,0.3)$)
    -- node[slbl, above] {reset: drop instance} ($(i2.north |- pool.north)+(0,0.3)$)
    -- (i2.north |- pool.north);

    \draw[hit] ([xshift=-1.05cm]reg.west) -- node[slbl, above] {cold} (reg.west);
    \draw[hit] ([xshift=-1.05cm]t1.west) -- node[slbl, above] {warm} (t1.west);
    \draw[hit] ([xshift=-1.05cm]t2.west) -- node[slbl, above] {hot} (t2.west);
  \end{tikzpicture}
  \caption{The cold, warm, and hot paths of an invocation on a node.}
  \label{fig:exec-path}
\end{figure}
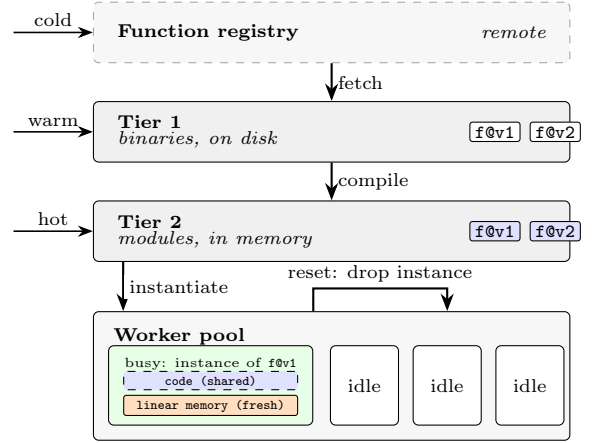

Each node keeps a pool of pre-warmed \emph{workers} and serves every invocation
it admits on one of them. The size of the pool bounds how many invocations the
node executes concurrently.
Before an instance can be created, though, the code of the function must be
available locally, and \fig{exec-path} shows the path an invocation follows.
A node retrieves the function binary from the function registry, which stores
the available functions as versioned WASM modules and is backed by MinIO\footnote{\url{https://www.min.io}}, an
S3-compatible object store, only on the first invocation of a given version.
To amortize this cost across invocations, the pool is backed by a two-tier
cache: the first tier stores the raw binaries retrieved from the registry, the
second the compiled modules in memory. A warm invocation thus only pays for
compilation, and a hot one bypasses both retrieval and compilation, executing
within a small constant factor of native code.
A worker serves an invocation by instantiating a cached module: the compiled
code is not copied but shared across concurrent instances, so that only the
linear memory, which every instance gets fresh and for itself, grows with
concurrency. That linear memory is also the channel through which the function
reaches the application state: the guest serializes the parameters of an
operation into its buffer and passes a pointer to a host function, which reads
them, dispatches the operation to the state management functionalities, and
writes the result back into the same buffer. The instance is discarded once the
execution ends, and the worker returns clean to the pool.

\subsection{State Management}
\label{sec:impl:state}

Among the data stores that WASP supports, Ermes uses Redis\footnote{\url{https://redis.io}} exclusively. As an
in-memory store, it keeps every state access off the disk, in line with the
platform's goal of minimizing state-access latency (\s{sys_overview}).
The node persists collections and their metadata in Redis, organized into three
keyspaces: \texttt{meta:} stores collection metadata as Redis hashes,
\texttt{data:} stores the application records, and \texttt{info:} stores
protocol state such as sequence numbers. Operations that must be atomic, such as
the last-writer-wins comparison-and-write and the commit of a batch, are
implemented as Lua scripts executed directly in Redis, with a retry loop under
randomized exponential backoff to absorb transient conflicts. On top of the
store, each node keeps an in-memory LRU cache that bounds the number of
collections it holds and drives eviction, and a \emph{partial view} that
materializes the reachability, traffic, and capacity information of its direct
neighbors.
The metadata-only index that backs collection resolution is built with
RediSearch: queries are answered with \texttt{FT.SEARCH} over the \texttt{meta:}
keyspace, and a user-defined metadata field introduced at runtime triggers an
\texttt{FT.ALTER} that extends the index schema without interrupting service.
This index is the concrete form of the reachability information that both
resolution and scheduling consult; because it records only properties and the
direction toward a replica, never the records, collections can be discovered
without replicating their data and can move without changing how functions refer
to them.

Access is exposed to the execution functionality through a uniform interface
that hides whether a collection resides locally or on a remote node. When a
function issues a \texttt{get} or a \texttt{put}, the host function of
\s{func:mng} dispatches the operation to this interface, which serves it from
the local store when a replica is present, and otherwise forwards it over HTTP
to the node that holds one. Before committing a write to a sequentially
consistent collection, the node validates the sequence number stored in the
\texttt{info:} keyspace, enforcing the ordering the leader establishes.

Placement is implemented as the finite-state machine of \s{state:mng}. Capacity
reports are exchanged over a dedicated HTTP endpoint, while per-neighbor,
per-collection demand is accumulated in penalized circular buffers held in the
partial view: the buffers retain the reports of a bounded window of past epochs
and weigh them by age, so that recent demand dominates the decision. In the
decision phase, the node evaluates its collections concurrently, one goroutine
each, and collects the outcomes through a channel; the resulting transfers, and
the migrations that hierarchical eviction adds when the cache occupancy exceeds
the safe level, are carried out over HTTP.

Replication drives the up-and-down protocol over the same HTTP transport. A node
maintains one outbound queue per neighbor, drained by a dedicated goroutine that
sends a single message and waits for its acknowledgment before the next, which
is what makes per-neighbor delivery ordered and keeps commits off the network.
The two migration safeguards each add a per-destination structure. On the
transferring node, a queue holds the updates postponed toward the destination,
shipped in order once the transfer is acknowledged and replayed on the local
replica if it is refused. On a node forwarding an update for a collection it
does not hold, a completion counter, decremented as each outgoing copy is
delivered, re-applies the update locally after the last delivery, in case a
replica has meanwhile arrived.
\section{Evaluation}
\label{sec:eval}

This section presents the experimental evaluation of Ermes.
The central promise of the platform is to reduce the latency a client
perceives when a function accesses application state. Ermes brings the state
close to where the computation runs and keeps it there as demand shifts,
rather than fetching that state from a remote data store.

Whereas our prior work~\cite{ermesTheory} studied the placement and
scheduling algorithm in simulation, here we evaluate the fully implemented
platform, and the latencies we report are those a client experiences end-to-end
from the running system.

We organize the evaluation around four research questions (RQ), each
isolating a distinct aspect of this promise:

\fakeparagraph{\textbf{RQ1}}
How does Ermes' state management reduce the latency a client observes?

\fakeparagraph{\textbf{RQ2}}
How does the client-perceived latency of Ermes evolve as the workload
submitted to the system grows?

\fakeparagraph{\textbf{RQ3}}
How effectively does Ermes adapt to client dynamics?

\fakeparagraph{\textbf{RQ4}}
How does Ermes compare, in client-perceived latency, against the cloud-based
deployments that represent current practice?

These questions follow the logic of the problem Ermes addresses. Stateless
FaaS platforms place computation close to the client but leave the state in
remote storage, so every state access pays a remote round-trip. RQ1 tests, in
isolation, whether relocating and replicating state actually shortens the path
an access travels, and thus the latency a client observes, and what each
consistency guarantee costs. RQ2 then asks whether this benefit survives at
scale, since a realistic deployment serves many clients over a large body of
state, with functions that differ in how much state they access. RQ3 asks
whether the benefit holds as clients evolve rather than staying fixed: we assess
how quickly the system converges for static clients and how well it follows
mobile ones, since a platform that could not track moving clients would not fit
the continuum. RQ4 finally places
these results in perspective, quantifying the gain of Ermes over the
cloud-based deployments that represent current practice, both at scale and
under client dynamics.

To answer RQ4 we compare Ermes against two cloud-based baselines that represent
current practice.
In the \emph{Cloud Only} (CO)\cite{openWhisk, openFaas} configuration, the
classic serverless arrangement, both the functions and the application state
reside in the cloud: an invocation issued at the edge is forwarded to the
cloud, executes there with its state local, and returns, so the client pays a
single round-trip to the cloud per invocation, regardless of how much state the
function accesses. In the \emph{Cloud Data} (CD)\footnote{\url{https://aws.amazon.com/lambda/edge/}} configuration instead, the application
state is pinned to the cloud, while functions are scheduled at the edge when possible, close
to the invoking client. Computation is thus brought close to the client, as in
Ermes, but the state is not, since edge nodes hold no state and act only as
executors, so every state access incurs a remote round-trip to the cloud.
We deliberately do not compare against another platform: to the best of our
knowledge, no complete stateful serverless platform for the edge-to-cloud
continuum with dynamic, per-collection replication exists to serve as a direct
competitor, so these two configurations are the most faithful references
available.

Before answering these questions, we fixed the execution substrate. Ermes
builds on WASP (\s{func:mng}), which supports several WebAssembly engines and
execution models; a detailed comparison among them, across execution time,
memory footprint and stability, and scalability under concurrent load, is
reported in the WASP paper~\cite{wasp}. Guided by that comparison, we adopt
Wasmtime with just-in-time compilation together with our hybrid caching
mechanism: this combination achieves near-native warm-start performance with a
stable memory footprint and full deployment portability, while avoiding the
architecture-specific binaries that ahead-of-time engines require.
Ermes holds state in memory, in Redis (\s{impl:state}), so the latencies we
report reflect the cost of locating, replicating, and coordinating state across
the hierarchy, not disk I/O.

The remainder of this section is organized as follows.
\s{eval:setup} details the experimental setup.
\s{eval:latency} characterizes the latency impact of state management (RQ1).
\s{eval:scalability} studies scalability under a growing workload and its
comparison against the baselines (RQ2 and RQ4).
\s{eval:adaptation} evaluates adaptation to static and mobile clients, again
against the baselines (RQ3 and RQ4).
\s{eval:discussion} summarizes the findings.

\subsection{Experimental Setup}
\label{sec:eval:setup}
We deployed Ermes on a virtualized infrastructure managed by Proxmox Virtual
Environment on an Intel Xeon Gold 6418H server.
We emulated a three-tier hierarchy consistent with the model of
\s{sys_overview}: one Cloud Node (\texttt{IT}, 8~vCPUs, 8~GB RAM), two Mid
Nodes (\texttt{IT/Rome} and \texttt{IT/Milan}, 4~vCPUs, 4~GB RAM each), and
five Edge Nodes (\texttt{IT/Rome/Edge1-2} and \texttt{IT/Milan/Edge1-3},
2~vCPUs, 2~GB RAM each), all running Ubuntu 24.04 LTS.
The centralized components of the Ermes platform (Group Token Provider and
Function Registry) are isolated on a dedicated virtual machine mirroring the
Cloud Node specifications.
Although the virtual machines are physically co-located, we used the Linux
Traffic Control (\texttt{tc}) subsystem with the NetEm scheduler to inject
realistic, heterogeneous RTT delays on each virtual network interface,
emulating the wide-area network characteristics of a geo-distributed topology.
The injected round-trip delays range from 5~ms to 20~ms per link, with the
shorter delays on the edge-to-fog links and the longer ones on the
fog-to-cloud links.
In all experiments, clients communicate exclusively with edge nodes and have
negligible latency toward them, while incurring the injected delays toward the
rest of the infrastructure.

The workload consists of clients that invoke functions reading and writing the
records of collections. The number of collections depends on the experiment and
reaches up to a thousand in the scalability study (\s{eval:scalability}).

\begin{figure}[tbp]
  \centering
  \includegraphics[width=\linewidth]{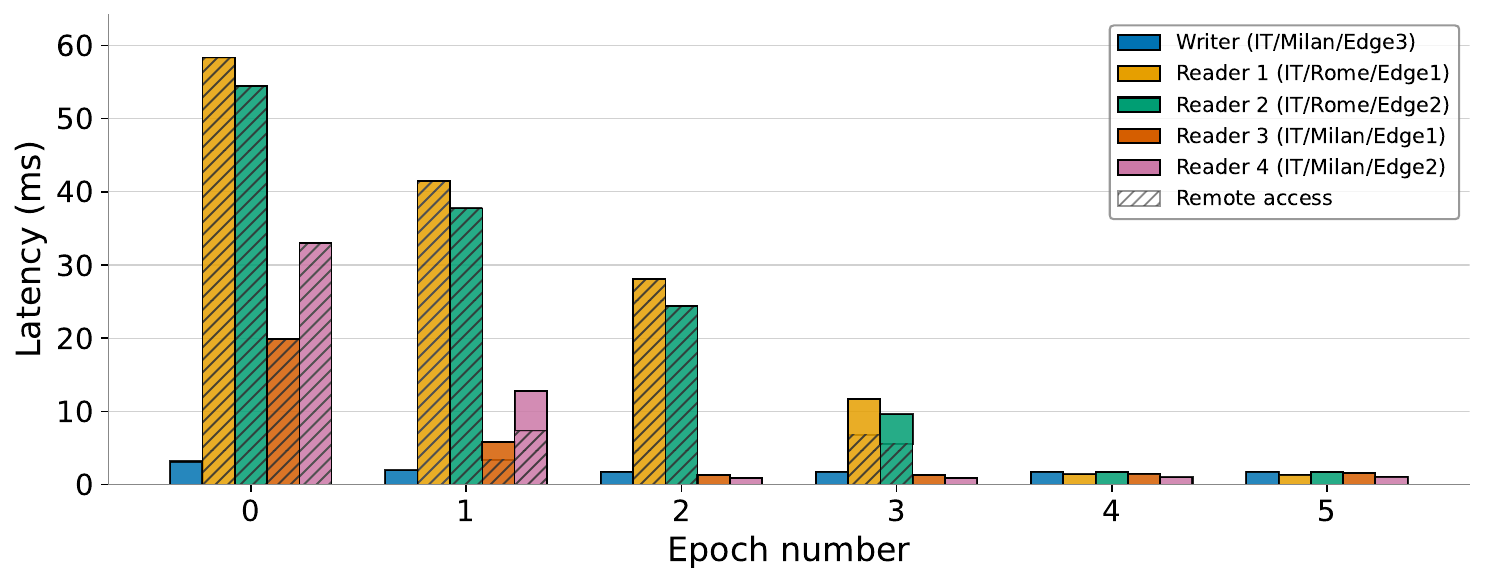}
  \caption{Per-epoch latency of five clients, one issuing writes and four
    issuing reads, each from a distinct edge node.}
  \label{fig:latency-consistency}
\end{figure}

\begin{figure}[tbp]
  \centering
  \includegraphics[width=\linewidth]{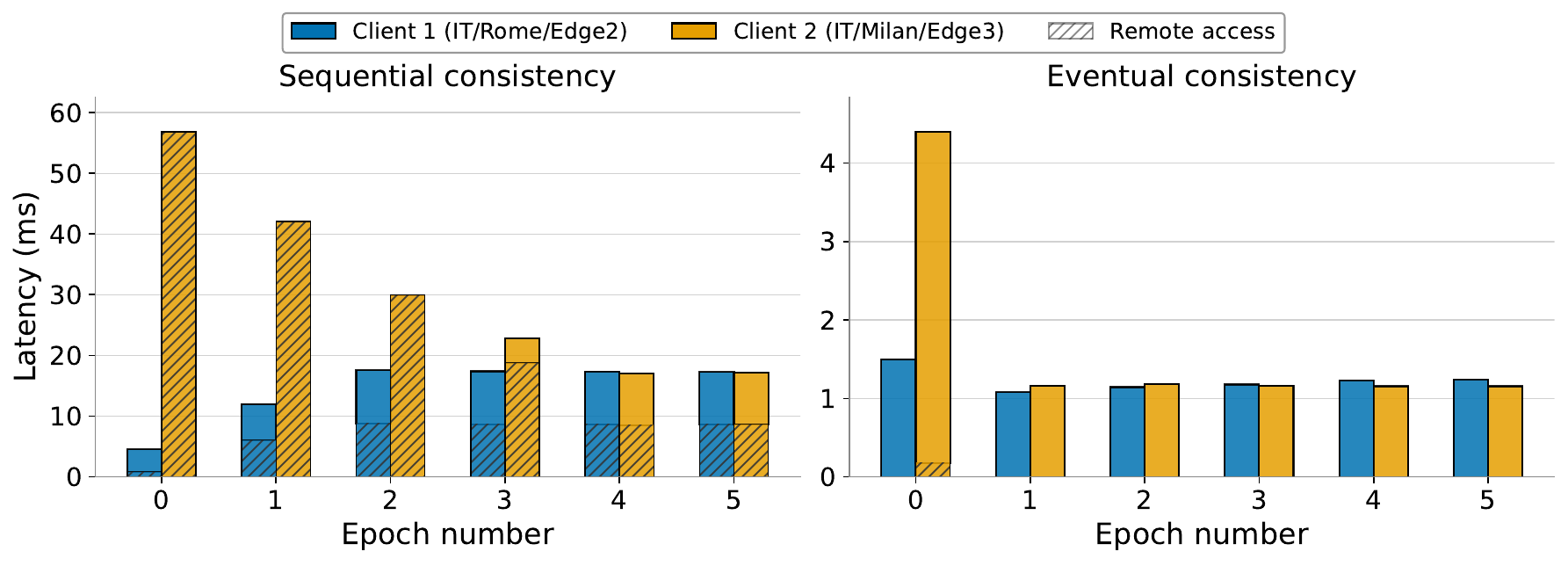}
  \caption{Per-epoch latency of two clients issuing reads and writes from
    opposite edges.}
  \label{fig:latency-consistency-comparison}
\end{figure}

\begin{figure*}[tbp]
  \centering
  \includegraphics[width=\textwidth]{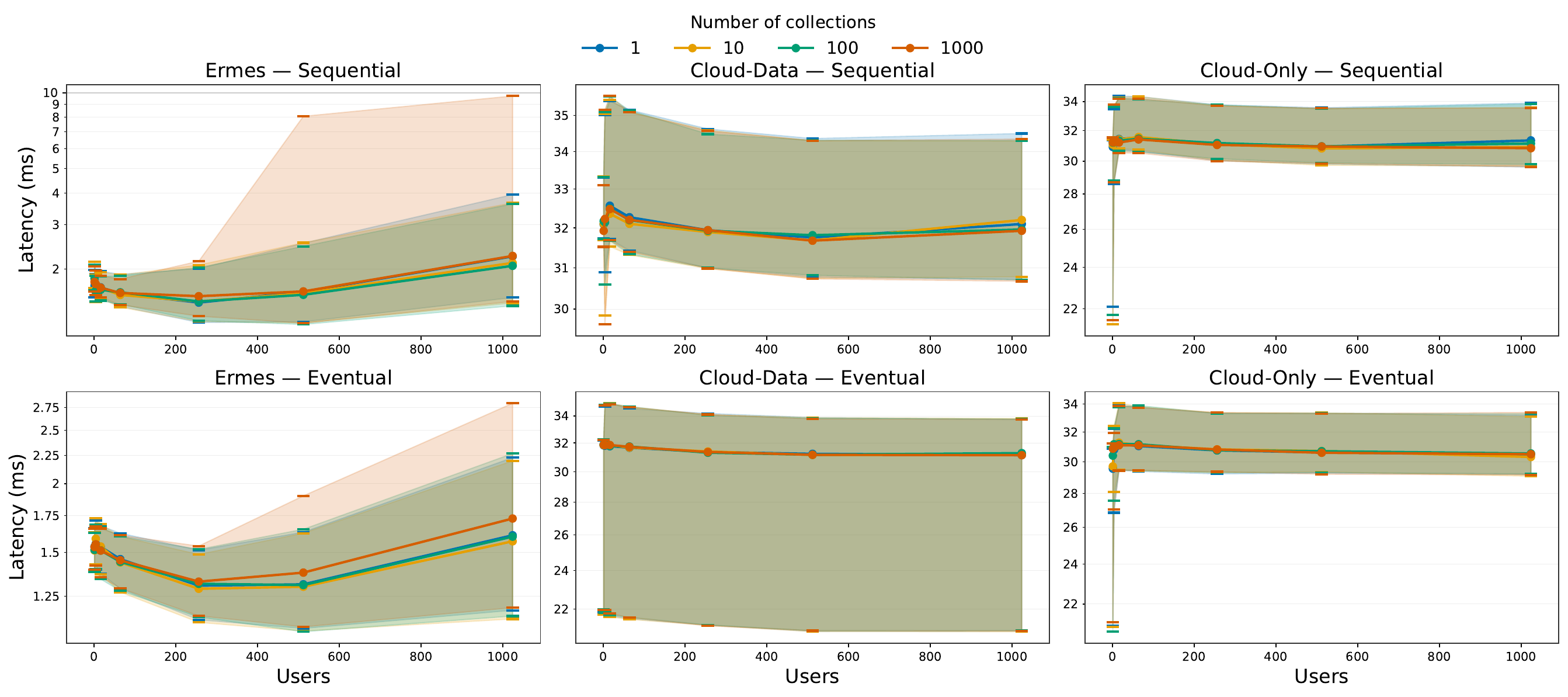}
  \caption{Steady-state latency versus concurrency and state cardinality.}
  \label{fig:scalability-all}
\end{figure*}

\subsection{State Management and Perceived Latency (RQ1)}
\label{sec:eval:latency}

To answer RQ1, we take the perspective of the individual client, reporting the
latency each one perceives as its accesses move from remote to local. We evaluate
two aspects of state management: how many epochs the placement mechanism takes
to turn a client's remote accesses into local ones, and the latency each client
observes at steady state under each of the two consistency policies.

We first isolate the reactivity of placement. Starting from a single-node
placement, we measure how quickly the framework detects a many-to-one read
pattern and provisions follower replicas accordingly, without degrading write
throughput.
One client continuously issues write invocations from \texttt{IT/Milan/Edge3},
while four clients at the remaining edge nodes issue read invocations against
the same collection.
Since placement evaluates each collection independently, the behavior observed
for a single collection generalizes to the multi-collection case: with $N$
independent collections, the system runs $N$ parallel placement decisions,
each following the same logic. The only source of inter-collection
interference is memory contention, which arises when the aggregate working set
exceeds a node's capacity and forces the migration of collections to the
parent node. We show in \s{eval:scalability} that placement handles this case
effectively, preserving quality of service under memory pressure.

\fig{latency-consistency} shows the latency evolution for each client over the
seven epochs (0 to 6) the scenario spans, distinguishing accesses served
locally from those that still incur at least one remote fetch.
Initially, the collection resides on the writer's node and the four readers
access it remotely, incurring high read latency: reads start between $\approx
  20$ and $\approx 60$~ms, depending on the reader's distance from the
collection. Placement detects the many-to-one read pattern and provisions four
follower replicas in parallel, one per reader node, without interrupting
ongoing writes.
Crucially, the writer's latency remains unaffected despite the additional
write traffic required to synchronize the four new replicas: it starts at
$\approx 2$~ms and settles around $\approx 1$~ms from epoch~2. The
read latency of the four readers likewise drops to $\approx 1$~ms once a local
follower is available, confirming that the framework manages multiple
concurrent replications without introducing bottlenecks.
The epoch at which each reader transitions from remote to local access varies
with its position in the topology: the two readers co-located with the
writer's Fog Node (\texttt{IT/Milan/Edge1-2}) converge at epoch~2, while the
two on the distant Fog Node (\texttt{IT/Rome/Edge1-2}) converge only at
epoch~4. This ordering reflects the hop-by-hop nature of placement, which
migrates the collection first toward the readers that are closer in latency
and hops, and reaches the more distant subtree later.
The elevated first epoch after convergence for each client is the cost of the
first execution of a function on a node, which downloads the function binary
from the Function Registry.

We then isolate the latency each consistency policy imposes under concurrent,
bidirectional traffic. Keeping the same infrastructure, two clients issue
alternating read and write invocations against the same collection from
opposite ends of the topology: one from \texttt{IT/Milan/Edge1}, co-located
with the initial collection placement, and one from the distant
\texttt{IT/Rome/Edge2}. Whereas the previous experiment showed how state
management removes remote reads, this one exposes the cost of remote writes,
where the two policies diverge.

\fig{latency-consistency-comparison} shows the latency evolution per epoch.
Under sequential consistency, the client co-located
with the collection initially executes locally, while the remote client incurs
high write latency, as writes must reach the leader replica to enforce a total
order. The system then converges to a stable equilibrium in which both clients
settle around $\approx 18$~ms. This is the structural cost of sequential
consistency: a single leader serializes the updates, so write traffic cannot be
local for all nodes at once. The value is deliberately a worst case: the two
writers sit at opposite edges of the topology, so the leader is necessarily
remote from one of them and each write traverses a long cross-topology path. The hatched bar portions confirm that
read traffic remains local throughout, served by follower replicas.
Under eventual consistency, latency drops to $\approx
  1$~ms from the first epoch onward, as both reads and writes are served locally
and state synchronization propagates asynchronously, without blocking user
requests.

Together, the two experiments answer RQ1: state management turns remote
accesses into local ones within a few epochs, and the residual latency a
client perceives at steady state is dictated by the consistency policy,
ranging from the near-local cost of eventual consistency to the single-leader
floor of sequential consistency.

\subsection{Scalability under a Growing Workload (RQ2 and RQ4)}
\label{sec:eval:scalability}

This section answers RQ2 and, through the comparison, RQ4. To answer RQ2 we
quantify how the latency characterized above holds as the workload grows; to
answer RQ4 we measure Ermes against the CD and CO baselines defined above. All
experiments are measured at steady state, after placement has settled. We
evaluate scalability along two axes: the size of
the workload, measured by the number of concurrent clients and the number of
collections, and the data-intensity of individual functions, measured by the
number of collection accesses per invocation.

We first vary the size of the workload through a stress test over two
independent load parameters: the \emph{concurrency level} $N_{users}$, from 1
to 1024 simultaneous clients, and the \emph{state cardinality} $N_{col}$, from
1 to 1000 collections, all of which are actively accessed by clients. We
increase the number of client-collection associations accordingly.
Each panel of \fig{scalability-all} plots, for one configuration and
consistency policy, the median request latency against $N_{users}$, with one
curve per value of $N_{col}$; the shaded band around each curve is the
interquartile range (IQR) of the measured latency.

In the CD baseline, latency is dominated by the network RTT to remote storage
and remains largely insensitive to both $N_{users}$ and $N_{col}$, since edge
nodes act only as stateless executors. Under both consistency policies, latency
sits around $\approx 32$~ms, with a wider IQR for sequential consistency.
The gap in the IQR reflects how writes are committed: they are synchronous under
sequential consistency, blocking until the remote commit completes, whereas
under eventual consistency they return immediately after local buffering, so
the RTT penalty is structurally lower.

The CO baseline is numerically almost indistinguishable from CD: latency is
insensitive to both $N_{col}$ and $N_{users}$, and sits around
$\approx 31$~ms under both consistency policies. Unlike CD, the two policies
also share the same narrow IQR ($\approx 30$--$34$~ms). With computation and state both in
the cloud, every state operation executes locally there, so the consistency
policy adds no network cost, and both the median and the IQR are set entirely
by the single round-trip that carries each invocation to the cloud.

In Ermes, under sequential consistency, every write to a collection must be
serialized by its single leader replica. As the number of concurrent clients
grows, more geographically dispersed writers contend for that leader, which
placement then keeps at a higher tier to balance them, so each write traverses
additional links. Latency is therefore driven by concurrency and stays largely
insensitive to $N_{col}$, staying between $\approx 1$~ms and $\approx 2$~ms. Memory pressure at high concurrency surfaces
only as a widening of the IQR, without shifting this median trend.

Under eventual consistency, the multi-leader design lets every replica accept
writes locally, so latency stays low across almost the entire range: it grows
only mildly with concurrency, from $\approx 1.5$~ms with a single client to
$\approx 1.75$~ms with 1024 clients.

In both cases, having 1000 collections combined with more than $\approx 256$
clients results in a widening of the IQR (up to $\approx 10$~ms for sequential and $\approx 2.75$~ms for eventual): the collections no longer fit in the
memory of the edge nodes, so some are placed one tier higher and accessed
remotely when possible.

Across the whole range, both Ermes configurations remain far below the
CD baseline: even at its worst operating point, Ermes latency
($\approx 10$~ms) is under a third of the $\approx 32$~ms the CD baseline pays
regardless of load.

We then vary the data-intensity of individual functions. Fixing concurrency
and state cardinality at representative values ($N_{users} = 64$, $N_{col} =
  100$), we vary the number of state operations per invocation from $1$ to $10$,
comparing Ermes against the CD and CO baselines. Writes are buffered and committed at
the end of the execution, so only reads contribute to the measured latency;
the x-axis therefore reports the number of reads per invocation, which are
issued sequentially, with no batching.

\begin{figure}[t]
  \centering
  \includegraphics[width=0.9\linewidth]{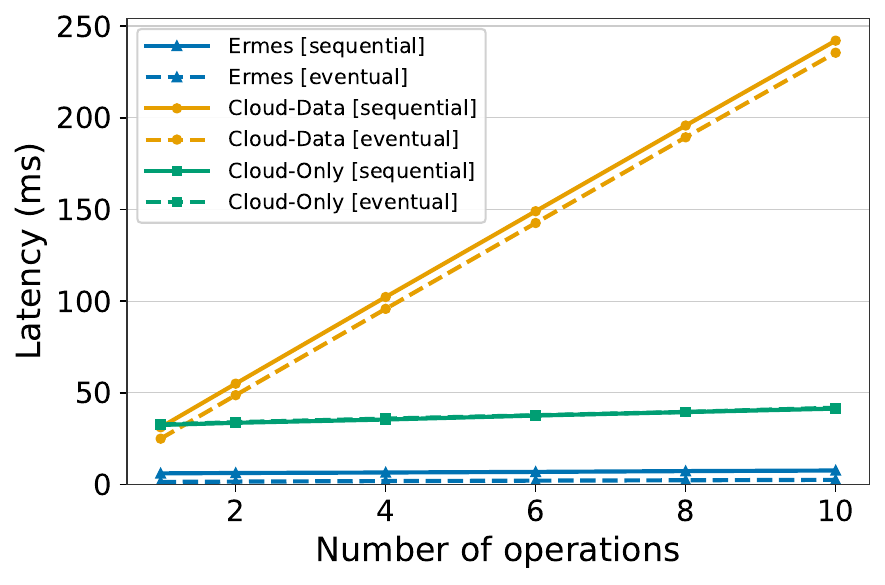}
  \caption{Steady-state latency versus data-intensity.}
  \label{fig:io-intensity-impact}
\end{figure}
\fig{io-intensity-impact} isolates the benefit of data locality.
In the CD baseline, latency grows linearly with the number of reads, from
$\approx 32$~ms at a single read to $\approx 301$~ms at ten: computation runs
at the edge but the state is remote, so each read is a separate round-trip up
to the cloud, and $T_{exec} \approx N_{reads} \times RTT_{cloud}$. Moving
computation to the edge without its state is thus actively harmful for
data-intensive functions.
The CO baseline, in contrast, is nearly flat, at $\approx 31$--$42$~ms:
functions execute in the cloud, co-located with the state, so reads are local
and the cost is dominated by the single round-trip that carries the invocation
to the cloud and back. The bottleneck is link traversal, not data access, so
the number of reads barely matters.
In Ermes, the profile is flat like CO, but an order of magnitude lower, at
$\approx 1.4$--$3$~ms across both policies, because the collection has been
replicated to the edge node serving the client, co-locating computation and
state at the edge, so every read is a local lookup with no round-trip at all.
The comparison highlights that removing the per-read network cost requires
co-locating computation and state, but only co-locating them at the edge, as
Ermes does, also eliminates the round-trip that keeps CO an order of magnitude
above it.

\subsection{Adaptation to Static and Mobile Clients (RQ3 and RQ4)}
\label{sec:eval:adaptation}

RQ2 characterized steady-state behavior; RQ3 asks how quickly and how
gracefully the system reaches and maintains it as clients evolve, and RQ4
compares this adaptation against the CD and CO baselines. Unlike RQ1, which
follows individual clients, here we take the aggregate view, reporting the
average latency across all clients and centering the analysis on the comparison
with the cloud-based baselines. We consider the two extremes of client dynamics:
static clients, for which placement must converge to a locality-optimal
configuration, and mobile clients, for which the state must continuously
migrate to follow the client across the edge tier.

We consider a heavy workload with high concurrency and large state scenario.
Both experiments fix the state at $N_{col} = 1000$ collections, with $N_{users} = 1024$ for the static scenario and $N_{users} = 64$ for the mobile one. With the number of nodes fixed, the client count governs how densely collections spread across the edge, so the two experiments cannot share it. The static experiment takes maximal concurrency. The mobility trace instead takes few clients: a denser population would keep collections replicated on every node, warming each destination in advance and hiding the migration cost, whereas a sparse one leaves destinations cold.

We first consider static clients, tracing the latency evolution across logical
epochs.

\begin{figure}[tbp]
  \centering
  \includegraphics[width=0.9\linewidth]{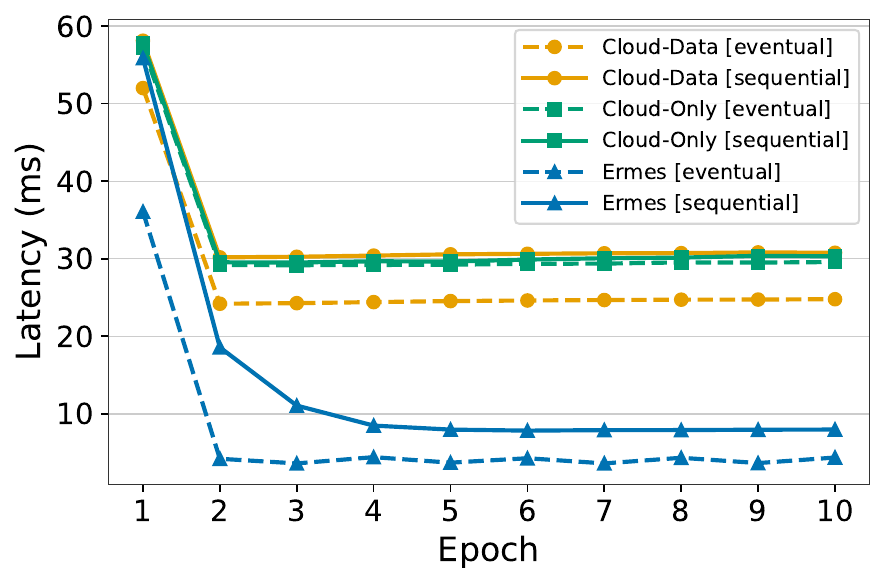}
  \caption{Latency across epochs for static clients.}
  \label{fig:temporal-analysis}
\end{figure}

\fig{temporal-analysis} presents this evolution. All baseline configurations
start high, between $\approx 55$ and $\approx 60$~ms, as the first epoch pays
the initial collection resolution, and then stay flat: CO, under both policies,
and CD under sequential consistency all settle at $\approx 30$~ms, dominated by
the round-trip to the cloud ($RTT_{cloud}$), since CO places both computation
and state there and the co-location constraint of sequential consistency forces
CD to do the same; CD under eventual consistency settles slightly lower, at
$\approx 25$~ms. None of the baselines improves further, since the state never
leaves the cloud. Ermes starts comparably high, at $\approx 55$~ms under
sequential and $\approx 45$~ms under eventual consistency, but, unlike the flat
baselines, converges downward within the first epochs as the collections
migrate to the edge nodes serving the clients: latency settles at $\approx
  8$~ms under sequential and $\approx 3$~ms under eventual consistency. Moving
the state to the edge thus proves faster than fetching it remotely even during
the initial convergence phase.

We then consider mobile clients, simulating a scenario in which all active
clients transition across the edge nodes in a round-robin fashion, changing
access point every five epochs.

\begin{figure}[tbp]
  \centering
  \includegraphics[width=0.9\linewidth]{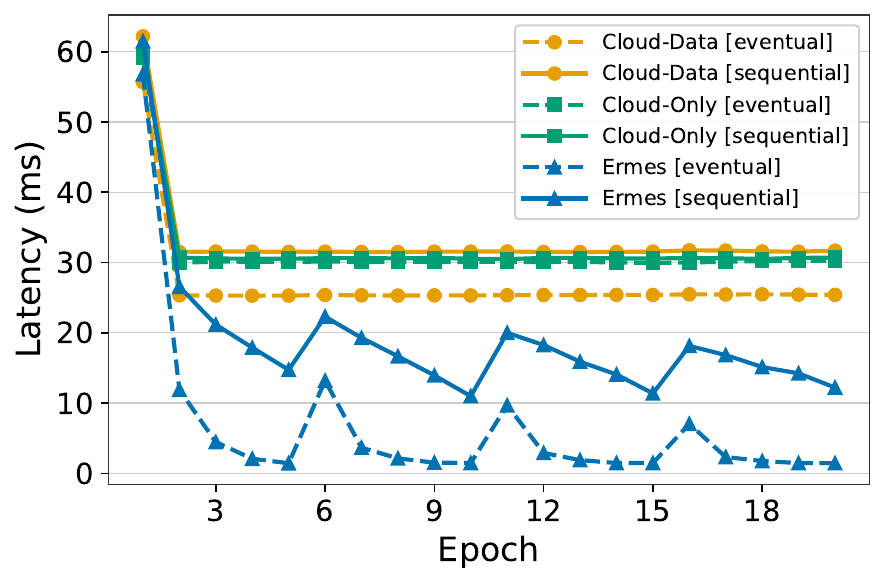}
  \caption{Latency across epochs for mobile clients.}
  \label{fig:mobility-trace}
\end{figure}

\fig{mobility-trace} compares the latency evolution. As in the static case,
the baselines are unaffected by mobility: all start around $\approx 60$~ms in
the first epoch and then settle at $\approx 30$~ms for CO (both policies) and
CD under sequential consistency, and at $\approx 25$~ms for CD under eventual
consistency. Ermes, in contrast, exhibits a characteristic sawtooth: starting
from $\approx 60$~ms, it converges to a low steady state punctuated by a spike
at each move, as placement reactively migrates the collection toward the
client's new location. Under eventual consistency, stable phases reach
$\approx 1$~ms and the spikes are sharp but bounded ($\approx 10$~ms, at
epochs~6, 11, and 16), with recovery in the following epoch. Under sequential
consistency, the steady state is higher ($\approx 10$~ms) and the spikes reach
$\approx 20$~ms, an overhead intrinsic to the single-leader protocol that
enforces total ordering; recovery is also slower. In both cases the latency
between moves stays well below the $\approx 25$--$30$~ms of the baselines,
confirming that Ermes preserves its locality advantage under continuous
mobility.

\subsection{Discussion}
\label{sec:eval:discussion}

Taken together, the experiments answer the four research questions for the
edge-to-cloud setting Ermes targets, in which clients invoke functions from the
edge. On the latency impact of state management (\textbf{RQ1}), placement turns remote
accesses into local ones within a few epochs, provisioning replicas without
perturbing ongoing writes, and the residual latency a client perceives is then
dictated by the consistency policy, near-local under eventual consistency and
bounded by the single-leader floor under sequential consistency. On
scalability (\textbf{RQ2}), Ermes sustains low and stable latency as
concurrency, state cardinality, and data-intensity grow, degrading gracefully
only under strict ordering or edge memory saturation. On adaptation to client
dynamics (\textbf{RQ3}), placement converges from the first epochs for static
clients and keeps pace with mobile ones, tracking each client as it roams. On
the comparison with cloud-based deployments (\textbf{RQ4}), Ermes outperforms
both the CD and CO baselines by a wide margin throughout; the comparison
further shows that co-locating computation and state cuts the per-access
network cost only when done at the edge, as in Ermes, and not in the cloud, as
in CO. The advantage is largest for read-intensive, eventually consistent, and mobile
workloads.
\section{Conclusion}
\label{sec:conclusion}
This paper presented Ermes, a stateful serverless platform that natively
integrates state management into the FaaS paradigm and jointly distributes
computation and application state across the edge-to-cloud continuum.
We evaluated Ermes on an emulated geo-distributed deployment, showing that
Ermes effectively brings state close to computation and keeps it there as
demand shifts, sustaining low client-perceived latency across a range of
workloads and consistency requirements and outperforming realistic deployment
solutions.

Several directions remain open.
Placement currently weighs the storage capacity of the nodes, and we plan to
extend it into a fully resource-aware policy that also accounts for the
compute available at each node.
On the consistency side, we intend to investigate protocols that adapt the
guarantee of a collection to its observed access pattern.
We also plan to strengthen the platform's fault tolerance: the multi-leader
design of eventually consistent collections, together with the cloud replica
that every collection retains, already makes recovery straightforward in most
scenarios, whereas sequential consistency additionally requires a leader
re-election mechanism to survive the failure of a leader, an orthogonal
concern we leave to future work.
Finally, we plan to evaluate Ermes on larger-scale physical infrastructures.
 
\section*{CRediT authorship contribution statement}

\textbf{Matteo Cenzato}: Conceptualization, Methodology, Software, Validation,
Visualization.
\textbf{Dario d'Abate}: Conceptualization, Methodology, Validation,
Visualization, Writing -- original draft, Writing -- review \& editing,
Supervision.
\textbf{Arianna Dragoni}: Conceptualization, Methodology, Validation,
Visualization, Writing -- original draft, Writing -- review \& editing,
Supervision.
\textbf{Giacomo Orsenigo}: Conceptualization, Methodology, Software,
Validation, Visualization.
\textbf{Luca Tosetti}: Conceptualization, Methodology, Software, Validation,
Visualization.
\textbf{Matteo Briscini}: Conceptualization, Methodology, Software,
Validation, Visualization.
\textbf{Alessandro Margara}: Conceptualization, Methodology, Validation,
Writing -- review \& editing, Supervision, Resources.

\section*{Data availability}
The source code of the Ermes platform, the experiment configurations and raw
results, and the code that regenerate all the figures in this paper are
openly available on Zenodo~\citep{artifact}.

\section*{Declaration of competing interest}
The authors declare that they have no known competing financial interests or
personal relationships that could have appeared to influence the work reported
in this paper.

\section*{Funding}
This research did not receive any specific grant from funding agencies in the
public, commercial, or not-for-profit sectors.

\section*{Declaration of generative AI and AI-assisted technologies in the manuscript preparation process}

During the preparation of this work, the authors used generative AI in order
to refine language editing and support software development;  all content was
reviewed and verified by the authors, who assume full responsibility for the
manuscript.

\bibliographystyle{elsarticle-num}

\end{document}